\documentclass{llncs}

\usepackage{mathptmx}
\usepackage{helvet}
\usepackage{courier}
\usepackage{multicol}
\usepackage[bottom]{footmisc}
\usepackage{amssymb}
\usepackage{rotating}

\usepackage{epsfig}
\usepackage{latexsym}
\usepackage{graphicx}
\usepackage{color}
\usepackage{url}
\usepackage{amsfonts}
\usepackage{pgf,pgfarrows,pgfnodes,pgfautomata,pgfheaps,pgfshade}
\usepackage{tikz}

\usetikzlibrary{arrows,decorations.pathmorphing,backgrounds,positioning,fit,petri}
\usepackage{etex}

\usepackage{stmaryrd}
\usepackage{amsmath}

\usepackage{bussproofs}

\usepackage{lineno}

\newcommand{\post}[1]{\mbox{$#1^{\bullet}$}}
\newcommand{\pre}[1]{\mbox{$^{\bullet}#1$}}

\newcommand{\fine}{{\mbox{ }\nolinebreak\hfill{$\Box$}}}

\newcommand{\deriv}[1]{{\mbox{${\:\stackrel{#1}{\longrightarrow}\:}$}}}

\newcommand{\nderiv}[1]{\nrightarrow}

\newcommand{\bigfrac}[2]{
\renewcommand{\arraystretch}{1.5}
\begin{array}{c}#1\\
\hline
#2
\end{array}}

\newcommand{\nil}{\mbox{\bf 0}}

\renewcommand{\mid}{\;\;\big|\;\;}

\newcommand{\nat}{{\mathbb N}}

\newcommand{\dom}{\mathit{dom}}

\begin{document}

 \pagestyle{headings}

\title{Place Bisimilarity is Decidable, Indeed!}
\author{Roberto Gorrieri\\
\institute{Dipartimento di Informatica --- Scienza e Ingegneria\\
Universit\`a di Bologna, \\Mura A. Zamboni, 7,
40127 Bologna, Italy}
\email{{\small roberto.gorrieri@unibo.it}}
}

\maketitle

\begin{abstract}
Place bisimilarity $\sim_p$ is a behavioral equivalence for finite Petri nets, originally proposed in \cite{ABS91},
that, differently from all the other behavioral relations proposed so far, is not defined 
over the markings of a finite net, rather over its 
places, which are finitely many. 
Place bisimilarity $\sim_p$ was claimed decidable in \cite{ABS91},
but its decidability was not really proved. 
We show that it is possible to decide $\sim_p$ with a simple algorithm, which essentially scans 
all the place relations (which are finitely many) to check whether they are place bisimulations.
We also show that $\sim_p$ does respect the intended causal semantics of Petri nets, as it is finer than 
causal-net bisimilarity \cite{Gor22}.
Moreover, we propose a slightly coarser variant, we call d-place bisimilarity $\sim_d$, that we conjecture to be the coarsest
equivalence, fully respecting causality and branching time (as it is finer than fully-concurrent bisimilarity \cite{BDKP91}), 
to be decidable on finite Petri nets.
Finally, two even coarser variants are discussed, namely i-place and i-d-place bisimilarities, which are still decidable, 
do preserve the concurrent behavior of Petri nets, but do not respect causality.
These results open the way towards formal verification (by equivalence checking) of distributed systems modeled by finite Petri nets.
\end{abstract}

%
\section{Introduction}
%

There is widespread agreement that a very natural equivalence relation for sequential, nondeterministic systems
is bisimulation equivalence \cite{Park81,Mil89}, defined over the semantic model of labeled transition systems \cite{Kel76}
(LTSs, for short). Bisimulation equivalence, which fully respects the branching structure of systems, 
is defined coinductively, has a nice fixpoint characterization, is easily decidable
for finite-state LTSs, even decidable on some classes of infinite state systems.

When considering distributed systems, one very natural semantic model is given by finite Petri nets 
(see, e.g., \cite{Pet81,Reisig,Gor17}),
in particular finite Place/Transition Petri nets \cite{DesRei98} (P/T nets, for short), a distributed model of computation 
that can represent a class of distributed systems where the number of sequential components can grow unboundedly.
Petri nets have been extensively used for the modeling and analysis of real distributed systems
(see, e.g., \cite{RR98b,Rei98} and the references therein).
However, (almost) all the behavioral equivalence relations that were defined on finite P/T nets (e.g., in the spectrum ranging from
{\em interleaving} bisimilarity to {\em fully concurrent} bisimilarity \cite{BDKP91}) 
are relations over the global states of the net, called {\em markings} (possibly decorated with some history information), 
which are usually infinitely many, and so, not surprisingly, these equivalence relations are all undecidable, as
observed by Jan\u{c}ar \cite{Jan95} and Esparza \cite{Esp98}. 
Indeed, by these observations, it seems impossible to algorithmically compare the behavior of finite P/T nets, in order
to verify the correctness of a distributed system by 
computing a behavioral equivalence relation between its
specification net and its implementation net, i.e., by means of {\em equivalence checking}, a well-known formal verification technique.
We want to contradict this idea by showing that this 
is possible, at least in principle.

As a matter of fact, one relevant exception exists: {\em place bisimilarity}, 
originating from an idea by Olderog
\cite{Old} (under the name of strong bisimilarity) and then refined by Autant, Belmesk and Schnoebelen \cite{ABS91}, 
which is an equivalence over markings, based on relations over the {\em finite set of net places}, 
rather than over the (possibly infinite) set
of net markings. This equivalence is very natural and intuitive: 
as a place can be interpreted as a sequential process type (and each token
in this place as an instance of a sequential process of that type), a place bisimulation
essentially states which kinds of sequential processes (composing the distributed system represented by the Petri net)
are to be considered as equivalent. 

Informally, a binary relation $R$ over the set $S$ of places  is a place bisimulation if for all markings $m_1$ and $m_2$
which are {\em bijectively} related via $R$ (denoted by $(m_1, m_2) \in R^\oplus$, where $R^\oplus$ is called the 
{\em additive closure} of $R$), if $m_1$ can perform transition $t_1$, reaching marking $m_1'$,
then $m_2$ can perform a transition $t_2$, reaching marking $m_2'$, 
such that the pre-sets of $t_1$ and $t_2$ are related by $R^\oplus$ (i.e., $(\pre{t_1}, \pre{t_2}) \in R^\oplus)$,
the label of $t_1$ and $t_2$ is the same, the post-sets of $t_1$ and $t_2$ are related by $R^\oplus$ 
(i.e., $(\post{t_1}, \post{t_2}) \in R^\oplus)$, and 
also for the reached markings we have $(m_1', m_2') \in R^\oplus$;
and symmetrically if $m_2$ moves first.  
Two markings $m_1$ and $m_2$ are place bisimilar, denoted by $m_1 \sim_p m_2$, if  a place bisimulation $R$ exists such
that $(m_1, m_2) \in R^\oplus$.  
Place bisimilarity does respect the expected causal behavior of Petri nets; as a matter of fact, 
we show that place bisimilarity is slightly finer than 
{\em causal-net bisimilarity} \cite{G15,Gor22} (or, equivalently, {\em structure preserving bisimilarity} \cite{G15}),
in turn slightly finer than {\em fully-concurrent bisimilarity} \cite{BDKP91} 
(or, equivalently, {\em history-preserving bisimilarity} \cite{RT88,vGG89,DDM89}).

Place bisimilarity is an equivalence relation, but its definition is not coinductive because the union of place bisimulations 
may be not a place bisimulation; so, in general, there is not one single largest place bisimulation, rather many maximal place bisimulations.
In fact, place bisimilarity is the relation on markings obtained by the union of the additive closure of each maximal place bisimulation.
As a consequence, the classic algorithms for computing the coinductive bisimulation equivalence 
on LTSs \cite{KS83,PT87} cannot be 
adapted to compute
place bisimilarity.

Place bisimilarity $\sim_p$ was claimed decidable in \cite{ABS91} and a polynomial-time 
algorithm was presented in \cite{AS92} to this aim. To be precise, the algorithm in \cite{AS92} is tailored
for {\em i-place bisimilarity} $\sim_i$ (cf. Section \ref{i-place-sec}), even if it is argued that this algorithm can be easily adapted for deciding also
place bisimilarity $\sim_p$.
However, (the variant of) the algorithm in \cite{AS92} does not characterize $\sim_p$, rather the 
subequivalence $\sim_{p_{eq}} \subseteq \sim_p$ 
corresponding to the only maximal place bisimulation which is also an equivalence relation 
(which is unique, indeed \cite{ABS91}); 
hence, the decidability of place bisimilarity was not proved.

The main contribution of this paper is to show that $\sim_p$ is decidable for finite P/T nets, indeed.
The proof idea is as follows. Note that a place relation $R \subseteq S \times S$ is finite if the set $S$ 
of places is finite; hence,
there are finitely many place relations for a finite net. We can list all these place relations, say $R_1, R_2, \ldots R_n$. 
It is possible to decide whether a place relation $R_i$ in this list is a place bisimulation by checking two {\em finite} conditions over 
a {\em finite} number of marking pairs: this is a non-obvious observation, as a place bisimulation requires
that the place bisimulation game holds for the infinitely many pairs $(m_1, m_2)$ belonging to $R_i^\oplus$. 
Hence, to decide whether 
$m_1 \sim_p m_2$, it is enough to check, for $i = $ $1, \ldots n$, whether $R_i$  is a place 
bisimulation and, in such a case, whether $(m_1, m_2) \in R_i^\oplus$.

Moreover, we propose a novel, slightly coarser variant, we call {\em d-place} bisimilarity and denote by $\sim_d$, 
that we conjecture to be the coarsest, {\em sensible}
(i.e., fully respecting causality and branching time)
equivalence to be decidable on finite P/T Petri nets (without silent moves). Essentially, a d-place bisimulation is a relation
on the set whose elements are the places {\em and} the empty marking, which is considered as an additional dummy place.
We prove that this weaker variant is still finer than {\em fully-concurrent} bisimilarity \cite{BDKP91}.
The decidability proof of d-place bisimilarity $\sim_d$ is very similar to that of place bisimilarity $\sim_p$.

Finally, we recall the coarsest variant of place bisimilarity proposed in \cite{AS92}, namely  
{\em i-place bisimilarity} $\sim_i$. An i-place bisimulation $R$ is a relation
such that $R^\oplus$ is an interleaving bisimulation. We argue that i-place bisimilarity $\sim_i$ is still decidable, 
by adapting the same proof technique
we provided for deciding place bisimilarity $\sim_p$. However,  $\sim_i$ is
not fully satisfactory: even if it respects the branching time, it does not respect causality. As a last minor contribution, 
we propose an even coarser decidable variant, called {\em i-d-place bisimilarity} $\sim_{id}$, that 
is finer than {\em step bisimilarity} $\sim_s$ (originally proposed in \cite{NT84}) and so it preserves the 
concurrent behavior of nets (but not causality). Indeed, $\sim_{id}$ is the coarsest
decidable equivalence defined so far on finite P/T Petri nets respecting concurrency and the branching time.

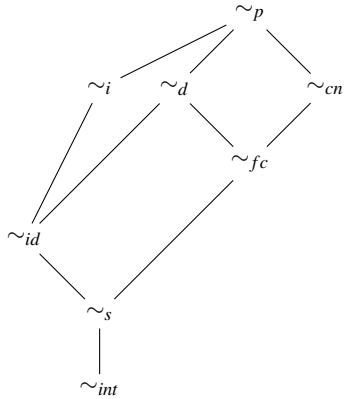
\begin{figure}[t]
\centering

\begin{tikzpicture}

\node (e1) at (6,15) {$\sim_p$};
\node (e2) at  (7,14) {$\sim_{cn}$};
\node (e3) at (5,14) {$\sim_d$};
\node (e4) at (4,14) {$\sim_i$};
\node (e5) at(3,12) {$\sim_{id}$};
\node (e6) at (6,13) {$\sim_{fc}$};
\node (e7) at (4,11) {$\sim_s$};
\node (e8) at (4,10) {$\sim_{int}$};

\draw  (e1) to (e2);
\draw  (e1) to (e3);
\draw  (e1) to (e4);
\draw  (e2) to (e6);
\draw  (e3) to (e6);
\draw  (e3) to (e5);
\draw  (e4) to (e5);
\draw  (e6) to (e7);
\draw  (e5) to (e7);
\draw  (e7) to (e8);

\end{tikzpicture}

\hrulefill

\caption{The diagram with the 8 behavioral equivalences studied in this paper}
\label{tabella-fig}
\end{figure}

Figure \ref{tabella-fig} summarizes the relations among the  eight different
behavioral equivalences we are studying in this paper, where the
top element is the most discriminating one (namely, place bisimilarity $\sim_p$) and the bottom element is the coarsest one
(namely, interleaving bisimilarity $\sim_{int}$). 
Note that the equivalences based on place relations 
(namely, place bisimilarity $\sim_p$, d-place bisimilarity $\sim_{d}$, i-place bisimilarity $\sim_i$ and
i-d-place bisimilarity $\sim_{id}$) are all decidable, while fully-concurrent bisimilarity
$\sim_{fc}$, step bisimilarity $\sim_s$ and interleaving bisimilarity $\sim_{int}$ are not; to complete the picture, the decidability of
causal-net bisimilarity $\sim_{cn}$ is an open problem.

The paper is organized as follows. Section \ref{def-sec} recalls the basic definitions about Petri nets, 
including their interleaving, concurrent and causal semantics.
Section \ref{place-sec} deals with place bisimilarity, shows that it is an equivalence relation, that it is not coinductive,
and, by means of examples, shows that $\sim_p$ is different from $\sim_{p_{eq}}$, i.e.,  
the only maximal place bisimulation which is also an equivalence relation (algorithmically characterized in \cite{AS92}).
Section \ref{place-cn-bis} proves that place bisimilarity implies causal-net bisimilarity.
Section \ref{decid-place-sec} proves that $\sim_p$ is decidable.
Section \ref{d-place-sec} introduces the coarser d-place bisimilarity $\sim_d$ and proves that it implies 
fully-concurrent bisimilarity and that it is still decidable.
Section \ref{i-place-sec} recalls i-place bisimilarity from \cite{AS92}, introduces the newly proposed coarser variant
i-d-place bisimilarity, shows that the latter implies step bisimilarity 
and hints that they are both decidable, too.
Finally, Section \ref{conc-sec} discusses the pros and cons of place bisimilarity, overviews related literature and hints some possible
future research.

%
\section{Petri Nets} \label{def-sec}
%

\begin{definition}\label{multiset}{\bf (Multiset)}\index{Multiset}
Let $\nat$ be the set of natural numbers. 
A {\em multiset} over a finite set $S$ is a function $m: S \rightarrow\nat$. 
The {\em support} set $dom(m)$ of $m$ is $\{ s \in S \mid m(s) \neq 0\}$. 
The set of all multisets 
over $S$,  denoted by ${\mathcal M}(S)$, is ranged over by $m$. 
We write $s \in m$ if $m(s)>0$.
The {\em multiplicity} of $s$ in $m$ is given by the number $m(s)$. The {\em size} of $m$, denoted by $|m|$,
is the number $\sum_{s\in S} m(s)$, i.e., the total number of its elements.
A multiset $m$ such 
that $dom(m) = \emptyset$ is called {\em empty} and is denoted by $\theta$.
We write $m \subseteq m'$ if $m(s) \leq m'(s)$ for all $s \in S$. 

{\em Multiset union} $\_ \oplus \_$ is defined as follows: $(m \oplus m')(s)$ $ = m(s) + m'(s)$; it 
is commutative, associative and has $\theta$ as neutral element. 
{\em Multiset difference} $\_ \ominus \_$ is defined as follows: 
$(m_1 \ominus m_2)(s) = max\{m_1(s) - m_2(s), 0\}$.
The {\em scalar product} of a number $j$ with $m$ is the multiset $j \cdot m$ defined as
$(j \cdot m)(s) = j \cdot (m(s))$. By $s_i$ we also denote the multiset with $s_i$ as its only element.
Hence, a multiset $m$ over $S = \{s_1, \ldots, s_n\}$
can be represented as $k_1\cdot s_{1} \oplus k_2 \cdot s_{2} \oplus \ldots \oplus k_n \cdot s_{n}$,
where $k_j = m(s_{j}) \geq 0$ for $j= 1, \ldots, n$.
\fine
\end{definition}

\begin{definition}\label{pt-net-def}{\bf (Place/Transition Petri net)}
A labeled {\em Place/Tran-sition} Petri net (P/T net for short) is a tuple $N = (S, A, T)$, where
\begin{itemize}
\item 
$S$ is the finite set of {\em places}, ranged over by $s$ (possibly indexed),
\item 
$A$ is the finite set of {\em labels}, ranged over by $\ell$ (possibly indexed), and
\item 
$T \subseteq {(\mathcal M}(S) \setminus \{\theta\}) \times A \times {\mathcal M}(S)$ 
is the finite set of {\em transitions}, 
ranged over by $t$ (possibly indexed).
\end{itemize}

Given a transition $t = (m, \ell, m')$,
we use the notation:
\begin{itemize}
\item  $\pre t$ to denote its {\em pre-set} $m$ (which cannot be an empty multiset) of tokens to be consumed;
\item $l(t)$ for its {\em label} $\ell$, and
\item $\post t$ to denote its {\em post-set} $m'$ of tokens to be produced.
\end{itemize} 

Hence, transition $t$ can be represented as $\pre t \deriv{l(t)} \post t$.
We also define the {\em flow function}
{\mbox  flow}$: (S \times T) \cup (T \times S) \rightarrow \nat$ as follows:
for all $s \in S$, for all $t \in T$,
{\mbox  flow}$(s,t) = \pre{t}(s)$ and {\mbox  flow}$(t,s) = \post{t}(s)$.
We will use $F$ to denote the {\em flow relation} 
$\{(x,y) \mid x,y \in S \cup T \, \wedge \, ${\mbox  flow}$(x,y) > 0\}$.
Finally, we define pre-sets and post-sets also for places as follows: $\pre s = \{t \in T \mid s \in \post t\}$
and $\post s = \{t \in T \mid s \in \pre t\}$. 
\fine
\end{definition}

Graphically, a place is represented by a little circle and a transition by a little box. These are 
connected by directed arcs, which
may be labeled by a positive integer, called the {\em weight}, to denote the number of tokens 
consumed (when the arc goes from a place to the transition) or produced (when the arc goes form the transition to a
place) by the execution of the transition;
if the number is omitted, then the weight default value of the arc is $1$.

\begin{definition}\label{net-system}{\bf (Marking, P/T net system)}
A multiset over $S$  is called a {\em marking}. Given a marking $m$ and a place $s$, 
we say that the place $s$ contains $m(s)$ {\em tokens}, graphically represented by $m(s)$ bullets
inside place $s$.
A {\em P/T net system} $N(m_0)$ is a tuple $(S, A, T, m_{0})$, where $(S,A, T)$ is a P/T net and $m_{0}$ is  
a marking over $S$, called
the {\em initial marking}. We also say that $N(m_0)$ is a {\em marked} net.
\fine
\end{definition}

%
\subsection{Sequential Semantics}\label{seql-sem-sec}
%

\begin{definition}\label{firing-system}{\bf (Enabling, firing sequence, transition sequence, reachable marking, safe net)}
Given a P/T net $N = (S, A, T)$, a transition $t $ is {\em enabled} at $m$, 
denoted by $m[t\rangle$, if $\pre t \subseteq m$. 
The execution (or {\em firing}) of  $t$ enabled at $m$ produces the marking $m' = (m \ominus  \pre t) \oplus \post t$. 
This is written $m[t\rangle m'$. 
A {\em firing sequence} starting at $m$ is defined inductively as follows:
\begin{itemize}
\item $m[\epsilon\rangle m$ is a firing sequence (where $\epsilon$ denotes an empty sequence of transitions) and
\item if $m[\sigma\rangle m'$ is a firing sequence and $m' [t\rangle m''$, then
$m [\sigma t\rangle m''$ is a firing sequence. 
\end{itemize}

The set of {\em reachable markings} from $m$ is 
$[m\rangle = \{m' \mid  \exists \sigma.$ $
m[\sigma\rangle m'\}$. 
The P/T net system $N = $ $(S, A, T, m_0)$ is  {\em safe} if  for each 
$m \in [m_0\rangle$ and for all $s \in S$, we have that $m(s) \leq 1$.
\fine
\end{definition}

Note that the reachable markings of a P/T net can be countably infinitely many when the net is not bounded, i.e.,
when the number of tokens on some places can grow unboundedly.

Now we recall a simple behavioral equivalence on P/T nets, defined directly over the markings of the net, which 
compares two markings with respect to their sequential behavior.

\begin{definition}\label{def-int-bis}{\bf (Interleaving Bisimulation)}
Let $N = (S, A, T)$ be a P/T net. 
An {\em interleaving bisimulation} is a relation
$R\subseteq {\mathcal M}(S) \times {\mathcal M}(S)$ such that if $(m_1, m_2) \in R$
then
\begin{itemize}
\item $\forall t_1$ such that  $m_1[t_1\rangle m'_1$, $\exists t_2$ such that $m_2[t_2\rangle m'_2$ 
with $l(t_1) = l(t_2)$ and $(m'_1, m'_2) \in R$,
\item $\forall t_2$ such that  $m_2[t_2\rangle m'_2$, $\exists t_1$ such that $m_1[t_1\rangle m'_1$ 
with $l(t_1) = l(t_2)$ and $(m'_1, m'_2) \in R$.
\end{itemize}

Two markings $m_1$ and $m_2$ are {\em interleaving bisimilar}, 
denoted by $m_1 \sim_{int} m_2$, if there exists an interleaving bisimulation $R$ such that $(m_1, m_2) \in R$.
\fine
\end{definition}

Interleaving bisimilarity was proved undecidable in \cite{Jan95} for P/T nets having at least two unbounded places, 
with a proof based on the comparison of two {\em sequential} P/T nets, 
where a P/T net is sequential if it does not offer any concurrent behavior. Hence, interleaving bisimulation equivalence is 
undecidable even for the subclass of sequential finite P/T nets. Esparza observed in \cite{Esp98} that all the non-interleaving 
bisimulation-based equivalences (in the spectrum ranging from interleaving bisimilarity to fully-concurrent bisimilarity \cite{BDKP91})
collapse to interleaving bisimilarity over sequential P/T nets. Hence, the proof in \cite{Jan95} applies to all these
non-interleaving bisimulation equivalences as well.

%
\subsection{Concurrent Semantics}\label{parallel-sem-sec}
%

Given a P/T net $N = (S, A, T)$ and a marking $m$, we say that two transitions $t_1, t_2 \in T$
are {\em concurrently enabled} at $m$ if $\pre t_1 \oplus \pre t_2 \subseteq m$. The {\em concurrent firing} of these
two transitions produces the marking $m' = (m \ominus (\pre t_1 \oplus \pre t_2)) \oplus (\post{t_1} \oplus \post{t_2})$.
We denote this fact by $m [\{t_1, t_2\}\rangle m'$.
It is also possible that the same transition is {\em self-concurrent} at some marking $m$, meaning that two or more 
occurrences of it are concurrently enabled at $m$. 

We can generalize the definition of concurrently enabled transitions to 
a finite, nonempty multiset $G$ over the set $T$, called a {\em step}.
A step $G: T \rightarrow \nat$ is enabled at marking $m$ if $\pre{G} \subseteq m$, where 
$\pre{G} =\bigoplus_{t \in T}G(t)\cdot \pre{t}$ and $G(t)$
denotes the number of occurrences of transition $t$ in the step $G$.
The execution of a step $G$ enabled at $m$ produces the marking $m' = (m \ominus \pre{G})\oplus \post{G}$, 
where $\post{G} = \bigoplus_{t \in T}G(t)\cdot\post{t}$. This is written $m[G\rangle m'$. 
We sometimes refer to
this as the {\em concurrent token game}, in opposition to the {\em sequential} token game of Definition \ref{firing-system}.
The label $l(G)$ of a step $G$ is the multiset $l(G):A \rightarrow \nat$ defined as follows:
$l(G)(a) = \sum_{t_i \in \dom(G). l(t_i) = a} G(t_i)$.

Now we define a notion of bisimulation based on the firing of steps, rather than of single transitions (as for interleaving bisimulation),
originally proposed in \cite{NT84}.

\begin{definition}\label{def-step-bis}{\bf (Step Bisimulation)}
Let $N = (S, A, T)$ be a P/T net. 
A {\em step bisimulation} is a relation
$R\subseteq {\mathcal M}(S) \times {\mathcal M}(S)$ such that if $(m_1, m_2) \in R$
then
\begin{itemize}
\item $\forall G_1$ such that  $m_1[G_1\rangle m'_1$, $\exists G_2$ s.t. $m_2[G_2\rangle m'_2$ 
with $l(G_1) = l(G_2)$ and $(m'_1, m'_2) \in R$,
\item $\forall G_2$ such that  $m_2[G_2\rangle m'_2$, $\exists G_1$ s.t. $m_1[G_1\rangle m'_1$ 
with $l(G_1) = l(G_2)$ and $(m'_1, m'_2) \in R$.
\end{itemize}

Two markings $m_1$ and $m_2$ are {\em step bisimilar} (or {\em step bisimulation equivalent}), 
denoted by $m_1 \sim_{s} m_2$, if there exists a step bisimulation $R$ such that $(m_1, m_2) \in R$.
\fine
\end{definition}

Of course, $\sim_{s}$ is finer than $\sim_{int}$; moreover,
also step bisimilarity is undecidable for P/T nets having at least two unbounded places \cite{Esp98}.

%
\subsection{Causality-based Semantics}\label{causal-sem-sec}
%

We outline some definitions, adapted from the literature
(cf., e.g., \cite{GR83,BD87,Old,G15,Gor22}).

\begin{definition}\label{acyc-def}{\bf (Acyclic net)}
A P/T net $N = (S, A, T)$ is
 {\em acyclic} if its flow relation $F$ is acyclic (i.e., $\not \exists x$ such that $x F^+ x$, 
 where $F^+$ is the transitive closure of $F$).
 \fine
\end{definition}

The causal semantics of a marked P/T net is defined by a class of particular acyclic safe nets, 
where places are not branched (hence they represent a single run) and all arcs have weight 1. 
This kind of net is called {\em causal net}. 
We use the name $C$ (possibly indexed) to denote a causal net, the set $B$ to denote its 
places (called {\em conditions}), the set $E$ to denote its transitions 
(called {\em events}), and
$L$ to denote its labels.

\begin{definition}\label{causalnet-def}{\bf (Causal net)}
A causal net is a finite marked net $C(\mathsf{m}_0) = (B,L, 
E,  \mathsf{m}_0)$ satisfying
the following conditions:
\begin{enumerate}
\item $C$ is acyclic;
\item $\forall b \in B \; \; | \pre{b} | \leq 1\, \wedge \, | \post{b} | \leq 1$ (i.e., the places are not branched);
\item  $ \forall b \in B \; \; \mathsf{m}_0(b)   =  \begin{cases}
 1 & \mbox{if $\; \pre{b} = \emptyset$}\\ 
  0  & \mbox{otherwise;}   
   \end{cases}$\\
\item $\forall e \in E \; \; \pre{e}(b) \leq 1 \, \wedge \, \post{e}(b) \leq 1$ for all $b \in B$ (i.e., all the arcs have weight $1$).
\end{enumerate}
We denote by $Min(C)$ the set $\mathsf{m}_0$, and by $Max(C)$ the set
$\{b \in B \mid \post{b} = \emptyset\}$.
\fine
\end{definition}

Note that any reachable marking of a causal net is a set, i.e., 
this net is {\em safe}; in fact, the initial marking is a set and, 
assuming by induction that a reachable marking $\mathsf{m}$ is a set and enables $e$, i.e., 
$\mathsf{m}[e\rangle \mathsf{m}'$,
then also
$\mathsf{m}' =  (\mathsf{m} \ominus \pre{e}) \oplus \post{e}$ is a set, 
as the net is acyclic and because
of the condition on the shape of the post-set of $e$ (weights can only be $1$).

 As the initial marking of a causal net is fixed by its shape (according to item $3$ of 
Definition \ref{causalnet-def}), in the following, in order to make the 
 notation lighter, we often omit the indication of the initial marking, 
 so that the causal 
 net $C(\mathsf{m}_0)$ is denoted by $C$.

\begin{definition}\label{trans-causal}{\bf (Moves of a causal net)}
Given two causal nets $C = (B, L, E,  \mathsf{m}_0)$
and $C' = (B', L, E',  \mathsf{m}_0)$, we say that $C$
moves in one step to $C'$ through $e$, denoted by
$C [e\rangle C'$, if $\; \pre{e} \subseteq Max(C)$, $E' = E \cup \{e\}$
and $B' = B \cup \post{e}$. 
\fine
\end{definition}

\begin{definition}\label{folding-def}{\bf (Folding and Process)}
A {\em folding} from a causal net $C = (B, L, E, \mathsf{m}_0)$ into a net system
$N(m_0) = (S, A, T, m_0)$ is a function $\rho: B \cup E \to S \cup T$, which is type-preserving, i.e., 
such that $\rho(B) \subseteq S$ and $\rho(E) \subseteq T$, satisfying the following:
\begin{itemize}
\item $L = A$ and $\mathsf{l}(e) = l(\rho(e))$ for all $e \in E$;
\item $\rho(\mathsf{m}_0) = m_0$, i.e., $m_0(s) = | \rho^{-1}(s) \cap \mathsf{m}_0 |$;
\item $\forall e \in E, \rho(\pre{e}) = \pre{\rho(e)}$, i.e., $\rho(\pre{e})(s) = | \rho^{-1}(s) \cap \pre{e} |$
for all $s \in S$;
\item $\forall e \in E, \, \rho(\post{e}) = \post{\rho(e)}$,  i.e., $\rho(\post{e})(s) = | \rho^{-1}(s) \cap \post{e} |$
for all $s \in S$.
\end{itemize}
A pair $(C, \rho)$, where $C$ is a causal net and $\rho$ a folding from  
$C$ to a net system $N(m_0)$, is a {\em process} of $N(m_0)$. 
\fine
\end{definition}

\begin{definition}\label{trans-process}{\bf (Moves of a process)}
Let $N(m_0) = (S, A, T, m_0)$ be a net system 
and let $(C_i, \rho_i)$, for $i = 1, 2$, be two processes of $N(m_0)$.
We say that $(C_1, \rho_1)$
moves in one step to $(C_2, \rho_2)$ through $e$, denoted by
$(C_1, \rho_1) \deriv{e} (C_2, \rho_2)$, if $C_1 [e\rangle C_2$
and $\rho_1 \subseteq \rho_2$.
\noindent
If $\pi_1 = (C_1, \rho_1)$ and $\pi_2 = (C_2, \rho_2)$, we denote
the move as $\pi_1 \deriv{e} \pi_2$.
\fine
\end{definition}

\begin{definition}\label{cn-bis-def}{\bf (Causal-net bisimulation)}
Let $N = (S, A, T)$ be a finite P/T net. A {\em causal-net bisimulation} 
is a relation $R$, composed of 
triples of the form $(\rho_1, C, \rho_2)$, where, for $i = 1, 2$, $(C, \rho_i)$ is a process 
of $N(m_{0_i})$ for some $m_{0_i}$,
such that if $(\rho_1, C, \rho_2) \in R$ then

\begin{itemize}
\item[$i)$] 
$\forall t_1, C', \rho_1'$ s.t.  $(C, \rho_1) \deriv{e} (C', \rho_1')$,
where $\rho_1'(e) = t_1$,
$\exists t_2, \rho_2'$ s.t.
$(C, \rho_2) \deriv{e} (C', \rho_2')$,
where $\rho_2'(e) = t_2$, and
$(\rho'_1, C', \rho'_2) \in R$;

\item[$ii)$] and
$\forall t_2, C', \rho_2'$ such that $(C, \rho_2) \deriv{e} (C', \rho_2')$,
where $\rho_2'(e) = t_2$,
$\exists t_1, \rho_1'$ such that
$(C, \rho_1) \deriv{e} (C', \rho_1')$,
where $\rho_1'(e) = t_1$, and
$(\rho'_1, C', \rho'_2) \in R$.

\end{itemize}

Two markings $m_{1}$ and $m_2$ of $N$ are cn-bisimilar (or cn-bisimulation equivalent), 
denoted by $m_{1} \sim_{cn} m_{2}$, 
if there exists a causal-net bisimulation $R$ containing a triple $(\rho^0_1, C^0, \rho^0_2)$, 
where $C^0$ contains no events and 
$\rho^0_i(Min( C^0))  = \rho^0_i(Max( C^0)) = m_i\;$ for $i = 1, 2$.
\fine
\end{definition}

Causal-net bisimilarity, which coincides with {\em structure-preserving bisimilarity} \cite{G15}, observes 
not only the events, but also the structure of the distributed state. The problem of deciding
causal-net bisimilarity over general P/T nets is open, as the observation of Esparza \cite{Esp98}
does not apply to causal-net bisimilarity.
It is slightly stronger than
{\em fully-concurrent bisimulation} (fc-bisimulation, for short) \cite{BDKP91} (as this equivalence observes only the events), whose definition
was inspired by previous notions of equivalence on other models of concurrency:
{\em history-preserving bisimulation}, originally defined in \cite{RT88} under the name of {\em behavior-structure bisimulation}, and 
then elaborated on in \cite{vGG89} (who called it by this name) and also independently defined in \cite{DDM89} 
(who called it by {\em mixed ordering bisimulation}). Its definition follows.

\begin{definition}\label{po-process-def}{\bf (Partial orders of events from a process)}
From a causal net $C = (B, L, E, \mathsf{m}_0)$,  we can 
extract the {\em partial order of its events}
$\mathsf{E}_{\mathsf{C}} = (E, \preceq)$,
where $e_1 \preceq e_2$ if there is a path in the net from $e_1$ to $e_2$, i.e., if $e_1 \mathsf{F}^* e_2$, where
$\mathsf{F}^*$ is the reflexive and transitive closure of
$\mathsf{F}$, which is the flow relation for $C$.

Given a process $\pi = (C, \rho)$, we denote $\preceq$ as $\leq_\pi$, 
i.e. given $e_1, e_2 \in E$, $e_1 \leq_\pi e_2$ if and only if $e_1 \preceq e_2$.

Two partial orders $(E_1, \preceq_1)$ and $(E_2, \preceq_2)$ are isomorphic 
if there is a label-preserving, order-preserving  map $g: E_1 \to E_2$, i.e., a bijection such that
$\mathsf{l}_1(e) = \mathsf{l}_2(g(e))$ and $e \preceq_1 e'$ if and only if $g(e) \preceq_2 g(e')$.
We also say that $g$ is an
{\em event isomorphism} between the causal nets 
$C_1$ and 
$C_2$ if it is an isomorphism between their associated 
partial orders $\mathsf{E}_{C_1}$
and $\mathsf{E}_{C_2}$.
\fine
\end{definition}

\begin{definition}\label{sfc-bis-def}{\bf (Fully-concurrent bisimulation)}
Given a finite P/T net $N = (S, A, T)$, a {\em fully-concurrent bisimulation}
is a relation $R$, composed of 
triples of the form $(\pi_1, g, \pi_2) $, where, for $i = 1,2$, 
$\pi_i = (C_i, \rho_i)$ is a process of $N(m_{0i})$ for some $m_{0i}$ and
$g$ is an event isomorphism between $\mathsf{E}_{C_1}$ and $\mathsf{E}_{C_2}$, such that  
if $(\pi_1, g, \pi_2) \in R$ then

\begin{itemize}
\item[$i)$] 
$\forall t_1, \pi_1'$ such that $\pi_1 \deriv{e_1} \pi_1'$ with $\rho_1'(e_1) = t_1$, $\exists t_2, \pi_2', g'$ such that

\begin{enumerate}
\item $\pi_2 \deriv{e_2} \pi_2'$ with $\rho_2'(e_2) = t_2$;
 \item $g' = g \cup \{(e_1, e_2)\}$, and finally,
\item $(\pi_1', g', \pi_2') \in R$;
\end{enumerate}

\item[$ii)$] and symmetrically, if $\pi_2$ moves first.
\end{itemize}

Two markings $m_1, m_2$ are fc-bisimilar,
denoted by $m_1 \sim_{fc} m_2$, if a fully-concurrent bisimulation R exists,
containing a triple $(\pi^0_1, \emptyset, \pi^0_2)$ where 
$\pi^0_i = (C^0_i, \rho^0_i)$ such that 
$C^0_i$ contains no events and 
$\rho^0_i(Min(C^0_i))  = \rho^0_i(Max(C^0_i))$ $ = m_i\;$ for $i = 1, 2$.
\fine

\end{definition}

%
\section{Place Bisimilarity} \label{place-sec}
%

We now present the place bisimulation idea, introduced in \cite{ABS91} as an 
improvement of a behavioral relation proposed by Olderog in \cite{Old} on safe 
nets (called {\em strong bisimulation}), which fails to induce an equivalence relation.
First, an auxiliary definition.

%
\subsection{Additive Closure and its Properties} \label{add-sec}
%

\begin{definition}\label{add-eq}{\bf (Additive closure)}
Given a P/T net $N = (S, A, T)$ and a {\em place relation} $R \subseteq S \times S$, we define a {\em marking relation}
$R^\oplus \, \subseteq \, {\mathcal M}(S) \times {\mathcal M}(S)$, called 
the {\em additive closure} of $R$,
as the least relation induced by the following axiom and rule.\\

$\begin{array}{lllllllllll}
 \bigfrac{}{(\theta, \theta) \in  R^\oplus} & \;   &   \; 
 \bigfrac{(s_1, s_2) \in R \;  (m_1, m_2) \in R^\oplus }{(s_1 \oplus m_1, s_2 \oplus m_2) \in  R^\oplus }  \\
\end{array}$
\\[-.2cm]
\fine
\end{definition}

Note that, by definition, two markings are related by $R^\oplus$ only if they have the same size; 
in fact, the axiom states that
the empty marking is related to itself, while the rule, assuming by induction 
that $m_1$ and $m_2$ have the same size, ensures that $s_1 \oplus m_1$ and
$s_2 \oplus m_2$ have the same size.

\begin{proposition}\label{fin-k-add}
For each relation $R \subseteq S \times S$,  if $(m_1, m_2) \in R^\oplus$, 
then $|m_1| = |m_2|$.
\fine
\end{proposition}

Note also that the membership $(m_1, m_2) \in R^\oplus$ may be proved in several different ways, depending on the chosen order of the elements
of the two markings and on the definition of $R$. For instance, if $R = \{(s_1, s_3),$ $(s_1, s_4),$ $(s_2, s_3), (s_2, s_4)\}$,
then $(s_1 \oplus s_2, s_3 \oplus s_4) \in R^\oplus$ can be proved by means of the pairs $(s_1, s_3)$ and $(s_2, s_4)$,
as well as by means of $(s_1, s_4), (s_2, s_3)$.

An alternative way to define that two markings $m_1$ and $m_2$
are related by $R^\oplus$ is to state that $m_1$ can be represented as $s_1 \oplus s_2 \oplus \ldots \oplus s_k$, 
$m_2$ can be represented as $s_1' \oplus s_2' \oplus \ldots \oplus s_k'$ and $(s_i, s_i') \in R$ for $i = 1, \ldots, k$. 
Therefore, it is easy to observe that if $(m_1, m_2) \in  R^\oplus$ and $m_1' \subseteq m_1$, then there exists $m_2' \subseteq m_2$
such that $(m_1', m_2') \in R^\oplus$ and $(m_1 \ominus m_1', m_2 \ominus m_2') \in R^\oplus$.

\begin{proposition}\label{add-prop1}\cite{Gor17b}
For each place relation $R \subseteq S \times S$, the following hold:
\begin{enumerate}
\item If $R$ is an equivalence relation, then $R^\oplus$ is an equivalence relation.
\item If $R_1 \subseteq R_2$, then $R_1^\oplus \subseteq R_2^\oplus$, i.e., the additive closure is monotone.
\item If $(m_1, m_2) \in R^\oplus$ and $(m_1', m_2') \in R^\oplus$,
then $(m_1 \oplus m_1', m_2 \oplus m_2') \in R^\oplus$, i.e., the additive closure is additive.
\item If $R$ is an equivalence, $(m_1 \oplus m_1',$ $ m_2 \oplus m_2') \in  R^\oplus$ 
and $(m_1, m_2) \in R^\oplus$,
then $(m_1', m_2') \in R^\oplus$, i.e., the additive closure of an equivalence place relation is subtractive.\\[-1cm]
\end{enumerate}
\fine
\end{proposition}

When $R$ is an equivalence relation, it is rather easy to check whether two markings are related by $R^\oplus$. An 
algorithm, described in \cite{Gor17b}, establishes whether an $R$-preserving bijection exists between 
the two markings $m_1$ and $m_2$ of equal size $k$ in $O(k^2)$ time. 
Another algorithm,
described in \cite{Gor20c}, 
checks whether $(m_1, m_2) \in R^\oplus$ in $O(n)$ time, where $n$ is the size of $S$.
However, these performant algorithms heavily rely on the fact that $R$ is an equivalence relation, hence also 
subtractive (case 4 of Proposition \ref{add-prop1}).
If $R$ is not an equivalence relation, which is typical  for place bisimulations, the naive algorithm for checking 
whether $(m_1, m_2) \in R^\oplus$ would simply consider 
$m_1$ represented as $s_1 \oplus s_2 \oplus \ldots \oplus s_k$, and then would scan all the possible permutations of 
$m_2$, each represented as $s'_1 \oplus s'_2 \oplus \ldots \oplus s'_k$, 
to check that $(s_i, s_i') \in R$ for $i = 1, \ldots, k$. Of course, this naive algorithm has worst-case complexity $O(k!)$.

\begin{example}\label{nsubtractive}
Consider $R = \{(s_1, s_3),$  $(s_1, s_4), (s_2, s_4)\}$, which is not an equivalence relation.
Suppose we want to check whether $(s_1 \oplus s_2, s_4 \oplus s_3) \in R^\oplus$.
If we start by matching $(s_1, s_4) \in R$, then we would fail because the residual $(s_2, s_3)$ does not belong to $R$.
However, if we permute the second marking to $s_3 \oplus s_4$, then we succeed because the required pairs
$(s_1, s_3)$ and $(s_2, s_4)$ are both in $R$.
\fine
\end{example}
 
Nonetheless, the problem of checking if $(m_1, m_2) \in R^\oplus$ has polynomial time complexity
because it can be considered an instance of
the problem of finding a perfect matching in a bipartite graph, 
where the nodes are the tokens in the 
two markings and the edges
are defined by the relation $R$.  
In fact, the definition of the bipartite graph takes $O(k^2)$ time (where $k = |m_1| = |m_2|$) and, then, 
the Hopcroft-Karp-Karzanov algorithm  \cite{HK73,Kar73} for computing the maximum matching has
worst-case time complexity $O(h\sqrt{k})$, where $h$ is the number of the edges in the bipartire graph ($h \leq k^2$) and
to check whether the maximum matching is perfect can be done simply by checking that the size of the matching equals the number of nodes in each partition, i.e., $k$.
Hence, in evaluating the complexity of the algorithm in Section \ref{decid-place-sec}, we assume that the complexity of 
checking whether $(m_1, m_2) \in R^\oplus$ is in $O(k^2 \sqrt{k})$.

A related problem is that of computing, given a marking $m_1$ of size $k$, the set of all the markings $m_2$ such that 
$(m_1, m_2) \in R^\oplus$. This problem can be solved with a worst-case time complexity of $O(n^k)$ because each of the $k$
tokens in $m_1$ can be related via $R$ to $n$ places at most.


\begin{proposition}\label{add-prop2}\cite{Gor17b}
For each  place relations $R, R_1, R_2 \subseteq S \times S$, the following hold:
\begin{enumerate}
\item $\emptyset^\oplus = \{(\theta, \theta)\}$, i.e., the additive closure of the empty place relation
yields a singleton marking relation, relating the empty marking to itself.
\item $(\mathcal{I}_S)^\oplus = \mathcal{I}_M$, i.e., the additive closure of the
identity relation on places $\mathcal{I}_S = \{(s, s) \mid s \in S\}$ yields the identity relation on markings
$\mathcal{I}_M = \{(m, m) \mid $ $ m \in  {\mathcal M}(S)\}$.
\item $(R^\oplus)^{-1} = (R^{-1})^\oplus$, i.e., the inverse of an additively closed relation $R$ equals the additive closure
of its inverse $R^{-1}$.
\item $(R_1 \circ R_2)^\oplus = (R_1^\oplus) \circ (R_2^\oplus)$, i.e., the additive closure of the composition of two 
place relations equals the compositions of their additive closures.\\[-1cm]
\end{enumerate}
\fine
\end{proposition}

%
\subsection{Place Bisimulation and its Properties} \label{place-sub-sec}
%

We are now ready to introduce place bisimulation, which is a non-interleaving behavioral 
relation defined over the net places.

\begin{definition}\label{def-place-bis}{\bf (Place Bisimulation)}
Let $N = (S, A, T)$ be a P/T net. 
A {\em place bisimulation} is a relation
$R\subseteq S \times S$ such that if $(m_1, m_2) \in R^\oplus$
then
\begin{itemize}
\item $\forall t_1$ such that  $m_1[t_1\rangle m'_1$, $\exists t_2$ such that $m_2[t_2\rangle m'_2$ 
with $(\pre{t_1}, \pre{t_2}) \in R^\oplus$, $l(t_1) = l(t_2)$,  $(\post{t_1}, \post{t_2}) \in R^\oplus$ and, moreover, 
$(m_1', m_2') \in R^\oplus$; and symmetrically,
\item $\forall t_2$ such that  $m_2[t_2\rangle m'_2$, $\exists t_1$ such that $m_1[t_1\rangle m'_1$ 
with $(\pre{t_1}, \pre{t_2}) \in R^\oplus$, $l(t_1) = l(t_2)$,  $(\post{t_1}, \post{t_2}) \in R^\oplus$ and, moreover, 
$(m_1', m_2') \in R^\oplus$.
\end{itemize}

Two markings $m_1$ and $m_2$ are  {\em place bisimilar}, denoted by
$m_1 \sim_p m_2$, if there exists a place bisimulation $R$ such that $(m_1, m_2) \in R^\oplus$.
\fine
\end{definition}
 
Note that a place relation 
$R\subseteq S \times S$ is a place bisimulation if for all $(m_1, m_2) \in R^\oplus$ the two place bisimulation conditions above hold. So, it seems that infinitely many verifications are necessary to establish whether $R$ is a place bisimulation. 
In Section \ref{decid-place-sec} we prove
that, actually, only finitely many verifications are necessary.

In order to prove that $\sim_p$ is an equivalence relation, we now list some useful properties of 
place bisimulation relations. 

\begin{proposition}\label{pt-prop-bis}
For each P/T net $N = (S, A, T)$, the following hold:
\begin{enumerate}
\item The identity relation ${\mathcal I}_S = \{ (s, s) \mid s \in S \}$ is a place bisimulation;
\item the inverse relation $R^{-1} = \{ (s', s) \mid (s, s') \in R\}$ of a place bisimulation $R$ is a place bisimulation;
\item the relational composition $R_1 \circ R_2 = \{ (s, s'') \mid $ $\exists s'. (s, s') \in R_1 \wedge (s', s'') \in R_2 \}$ of
two place bisimulations $R_1$ and $R_2$ is a place bisimulation.
\end{enumerate}

\proof
The proof is almost standard, due to Proposition \ref{add-prop2}.

(1) 
${\mathcal I}_S$ is a place bisimulation as for each $(m, m) \in {\mathcal I}_S^\oplus$
whatever transition $t$ the left (or right) marking $m$ performs a transition (say,  $m[t\rangle m'$), 
the right (or left) 
instance of $m$
in the pair does exactly the same transition 
$m[t\rangle m'$ and, of course, $(\pre{t}, \pre{t}) \in {\mathcal I}_S^\oplus$, 
          $(\post{t}, \post{t}) \in {\mathcal I}_S^\oplus$, $l(t) = l(t)$, $(m', m') \in 
          {\mathcal I}_S^\oplus$, by Proposition \ref{add-prop2}(2), 
          as required by the place bisimulation definition.

(2) Suppose $(m_2, m_1) \in (R^{-1})^\oplus$ and $m_2[t_2\rangle m_2'$. By Proposition \ref{add-prop2}(3)
$(m_2, m_1) \in (R^\oplus)^{-1}$ and so $(m_1, m_2) \in R^\oplus$. Since $R$ is a place bisimulation, the second item of the 
bisimulation game ensures that there exist $t_1$ and $m_1'$
such that $m_1 [t_1\rangle m_1'$, with $(\pre{t_1}, \pre{t_2}) \in R^\oplus$, $l(t_1) = l(t_2)$, $(\post{t_1}, \post{t_2}) \in R^\oplus$
and $(m_1', m_2') \in R^\oplus$. Summing up, if $(m_2, m_1) \in (R^{-1})^\oplus$,
to the move $m_2[t_2\rangle m_2'$, $m_1$ replies with the move $m_1 [t_1\rangle m_1'$, such that 
(by Proposition \ref{add-prop2}(3))
$(\pre{t_2}, \pre{t_1}) \in (R^{-1})^\oplus$, 
$l(t_2) = l(t_1)$, $(\post{t_2}, \post{t_1}) \in (R^{-1})^\oplus$
and $(m_2', m_1') \in (R^{-1})^\oplus$, as required. The case when $m_1$ 
moves first is symmetric and so omitted.

(3)  Suppose $(m, m'') \in (R_1 \circ R_2)^\oplus$ and $m [t_1\rangle m_1$.
By Proposition \ref{add-prop2}(4), we have that 
$(m, m'') \in R_1^\oplus \circ R_2^\oplus$, and so there exists $m'$ such that $(m, m') \in R_1^\oplus$
and $(m', m'') \in R_2^\oplus$.
As $(m, m') \in R_1^\oplus$ and $R_1$ is a place bisimulation, 
if $m [t_1\rangle m_1$, then there exist $t_2$ and $m_2$ such that  
$m' [t_2\rangle m_2$ with $(\pre{t_1}, \pre{t_2}) \in R_1^\oplus$, $l(t_1) = l(t_2)$, $(\post{t_1}, \post{t_2}) \in R_1^\oplus$
and $(m_1, m_2) \in R_1^\oplus$.
But as $(m', m'') \in R_2^\oplus$ and $R_2$ is a place bisimulation, we have also that there exist $t_3$ and $m_3$ 
such that  $m'' [t_3 \rangle m_3$ 
with $(\pre{t_2}, \pre{t_3}) \in R_2^\oplus$, $l(t_2) = l(t_3)$, $(\post{t_2}, \post{t_3}) \in R_2^\oplus$
and $(m_2, m_3) \in R_2^\oplus$. 
Summing up, for any pair $(m, m'') \in (R_1 \circ R_2)^\oplus$, if $m [t_1 \rangle m_1$, then there exist $t_3$ and $m_3$ such that 
$m'' [t_3 \rangle m_3$ and (by Proposition \ref{add-prop2}(4)) $(\pre{t_1}, \pre{t_3}) \in (R_1 \circ R_2)^\oplus$, 
$l(t_1) = l(t_3)$, $(\post{t_1}, \post{t_3}) \in (R_1 \circ R_2)^\oplus$
and $(m_1, m_3) \in (R_1 \circ R_2)^\oplus$, as required. The case when $m''$ 
moves first is symmetric and so omitted.
\fine
\end{proposition}

\begin{proposition}\label{place-bis-eq}\index{Bisimulation!equivalence}
For each P/T net $N = (S, A, T)$, relation $\sim_p \; \subseteq  \mathcal{M}(S) \times  \mathcal{M}(S)$ is an equivalence relation.
\proof
As the identity relation ${\mathcal I}_S$ is a place bisimulation by Proposition \ref{pt-prop-bis}(1), 
we have that ${\mathcal I}_S^\oplus \subseteq \; \sim_p$, and so $\sim_p$ is reflexive.
Symmetry derives from the following argument.
For any $(m, m') \in \; \sim_p$, there exists a place bisimulation $R$ such that $(m, m') \in R^\oplus$; 
by Proposition \ref{pt-prop-bis}(2), relation $R^{-1}$ is a place bisimulation, and  by Proposition \ref{add-prop2}(3)
we have that $(m', m) \in (R^{-1})^\oplus$; hence,
$(m', m) \in \; \sim_p$.
Transitivity also holds for $\sim_p$. Let $(m, m') \in \; \sim_p$ and $(m', m'') \in \; \sim_p$; hence, there 
exist two place bisimulations $R_1$ and $R_2$ such that $(m, m') \in R_1^\oplus$ and $(m', m'') \in R_2^\oplus$. By 
Proposition \ref{pt-prop-bis}(3),  $R_1 \circ R_2$ is a place bisimulation such that the pair $(m, m'') \in (R_1 \circ R_2)^\oplus$  
by Proposition \ref{add-prop2}(4); hence, $(m, m'') \in \; \sim_p$.
\fine
\end{proposition}

\begin{proposition}\label{place-int-prop}{\bf (Place bisimilarity is finer than interleaving bisimilarity)}
Given a P/T net $N = (S, A, T)$ and two markings $m_1, m_2$, if $m_1 \sim_p m_2$, then $m_1 \sim_{int} m_2$.
\proof
If $m_1 \sim_p m_2$, then there exists a place bisimulation $R$ such that $(m_1, m_2) \in R^\oplus$.
Note that if $R$ is a place bisimulation, then $R^\oplus$ is an interleaving bisimulation, and so $m_1 \sim_{int} m_2$.
\fine
\end{proposition}

The implication above is strict, i.e., it is possible to find a P/T net and two markings that are interleaving bisimilar 
but not place bisimilar. For instance, in the CCS \cite{Mil89} terminology, the marking for $a.b.\nil + b.a.\nil$ 
and that for $a.\nil | b.\nil$ (see, e.g., \cite{Gor17}).
It is also possible to show that place bisimilarity implies step bisimilarity \cite{NT84} (cf. the subsequent Example \ref{ex-step} and 
Theorem \ref{id>step-th}).
As a matter of fact, place bisimilarity is a truly concurrent behavioral equivalence, which
is slightly finer than {\em causal-net bisimilarity} \cite{G15,Gor22} (or, equivalently, 
{\em structure preserving bisimilarity} \cite{G15}), 
in turn
slightly finer than {\em fully-concurrent bisimilarity} \cite{BDKP91}.
The following examples illustrate these relationships.

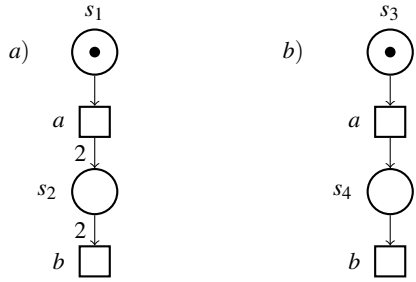
\begin{figure}[!t]
\centering
\begin{tikzpicture}[
every place/.style={draw,thick,inner sep=0pt,minimum size=6mm},
every transition/.style={draw,thick,inner sep=0pt,minimum size=4mm},
bend angle=30,
pre/.style={<-,shorten <=1pt,>=stealth,semithick},
post/.style={->,shorten >=1pt,>=stealth,semithick}
]
\def\eofigdist{3.3cm}
\def\eodist{0.4cm}
\def\eodisty{0.8cm}

\node (a) [label=left:$a)\qquad $]{};

\node (p1) [place, tokens=1]  [label=above:$s_1$] {};
\node (t1) [transition] [below =\eodist of p1,label=left:$a\;$] {};
\node (p2) [place] [below =\eodist of t1,label=left:$s_2\;$] {};
\node (t2) [transition] [below =\eodist of p2,label=left:$b\;$] {};

\draw  [->] (p1) to (t1);
\draw  [->] (t1) to node[auto,swap] {2} (p2);
\draw  [->] (p2) to node[auto,swap] {2} (t2);

  \node (b) [right={3.1cm} of a,label=left:$b)\quad$] {};

\node (p3) [place,tokens=1]  [right=\eofigdist of p1, label=above:$s_3$] {};
\node (t3) [transition] [below =\eodist of p3,label=left:$a\;$] {};
\node (p4) [place]  [below =\eodist of t3, label=left:$s_4\;$] {};
\node (t4) [transition] [below =\eodist of p4,label=left:$b\;$] {};

\draw  [->] (p3) to (t3);
\draw  [->] (t3) to (p4);
\draw  [->] (p4) to (t4);

\end{tikzpicture}
\caption{Two fully-concurrent bisimilar markings, which are not place bisimilar}
\label{net-fc-place}
\end{figure}

\begin{figure}[!t]
\centering
\begin{tikzpicture}[
every place/.style={draw,thick,inner sep=0pt,minimum size=6mm},
every transition/.style={draw,thick,inner sep=0pt,minimum size=4mm},
bend angle=30,
pre/.style={<-,shorten <=1pt,>=stealth,semithick},
post/.style={->,shorten >=1pt,>=stealth,semithick}
]
\def\eofigdist{3.3cm}
\def\eodist{0.4cm}
\def\eodisty{0.8cm}

\node (p1) [place,tokens=1]  [label=above:$s_1$] {};
\node (t1) [transition] [below =\eodist of p1,label=left:$a\;$] {};
\node (p2) [place] [below left =\eodist of t1,label=left:$s_2\;$] {};
\node (p3) [place] [below right =\eodist of t1,label=right:$\;s_3$] {};
\node (t2) [transition] [below right=\eodist of p2,label=left:$b\;$] {};
\node (p4) [place] [below =\eodist of t2,label=left:$s_4\;$] {};

\draw  [->] (p1) to (t1);
\draw  [->] (t1) to (p2);
\draw  [->] (t1) to (p3);
\draw  [->] (p2) to (t2);
\draw  [->] (p3) to (t2);
\draw  [->] (t2) to (p4);


\node (p5) [place,tokens=1]  [right=\eofigdist of p1, label=above:$s_5$] {};
\node (t4) [transition] [below =\eodist of p5,label=left:$a\;$] {};
\node (p6) [place]  [below =\eodist of t4, label=left:$s_6\;$] {};
\node (t5) [transition] [below =\eodist of p6,label=left:$b\;$] {};
\node (p7) [place]  [below =\eodist of t5,label=left:$s_7\;$] {};

\draw  [->] (p5) to (t4);
\draw  [->] (t4) to node[auto,swap] {2} (p6);
\draw  [->] (p6) to node[auto,swap] {2} (t5);
\draw  [->] (t5) to (p7);

\end{tikzpicture}
\caption{Two causal-net bisimilar markings, which are not place bisimilar}
\label{net-sp-place}
\end{figure}
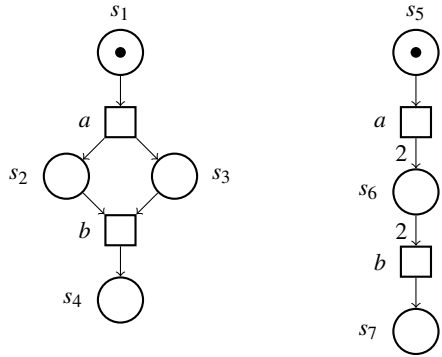

\begin{example}\label{ex-p-fc}
Consider the net in Figure \ref{net-fc-place}. Note that it is not possible to build a place bisimulation
relating $s_1$ and $s_3$  
because the post-sets of the
$a$-labeled transitions from $s_1$ and $s_3$ (i.e., $2 \cdot s_2$ and $s_4$, respectively) have different size.
Of course, the two markings $s_1$ and $s_3$ are not causal-net bisimilar, as they generate different causal nets.
However, $s_1$ and $s_3$ are fully-concurrent bisimilar \cite{BDKP91}, because both can perform the same partial 
order of events (i.e., $b$ caused by $a$). 
\fine
\end{example}

\begin{example}\label{ex-p>cn}
An intriguing example showing the difference between place bisimilarity and causal-net bisimilarity
is about the nets in Figure \ref{net-sp-place}. 
Of course, $s_1$ and $s_5$ are causal-net bisimilar, because both generate the same causal nets. 
However, no place bisimulation
relating $s_1$ and $s_5$  exists,
as this place relation would require the pair $(s_2, s_6)$, but $s_2$ and $s_6$ are not place bisimilar; 
in fact, $2 \cdot s_2$ is stuck, while $2 \cdot s_6$ can perform $b$. 
\fine
\end{example}

\noindent
By Definition \ref{def-place-bis}, place bisimilarity $\sim_p$ can be defined as follows:

$\sim_p = \bigcup \{ R^\oplus \mid R \mbox{ is a place bisimulation}\}.$

\noindent
By monotonicity of the additive closure (Proposition \ref{add-prop1}(2)), if $R_1 \subseteq R_2$, then
$R_1^\oplus \subseteq R_2^\oplus$. Hence, we can restrict our attention to maximal place bisimulations only:

$\sim_p = \bigcup \{ R^\oplus \mid R \mbox{ is a {\em maximal} place bisimulation}\}.$

\noindent
However, it is not true that 

$\sim_p = (\bigcup \{ R \mid R \mbox{ is a {\em maximal} place bisimulation}\})^\oplus$

\noindent 
because the union of place bisimulations may be not a place bisimulation (as already observed in \cite{ABS91}), 
so that its definition is not coinductive. We illustrate this fact by means of the following tiny example.

\begin{figure}[t]
\centering
\begin{tikzpicture}[
every place/.style={draw,thick,inner sep=0pt,minimum size=6mm},
every transition/.style={draw,thick,inner sep=0pt,minimum size=4mm},
bend angle=30,
pre/.style={<-,shorten <=1pt,>=stealth,semithick},
post/.style={->,shorten >=1pt,>=stealth,semithick}
]
\def\eofigdist{4cm}
\def\eodist{0.4cm}
\def\eodisty{0.8cm}

\node (p1) [place]  [label=left:$s_1\;$] {};
\node (t1) [transition] [below right=\eodist of p1,label=left:$a\;$] {};
\node (p2) [place] [right=\eodisty of p1,label=right:$\;s_2$] {};
\node (p3) [place] [below=\eodist of t1,label=left:$s_3\;$] {};

\draw  [->] (p1) to (t1);
\draw  [->] (p2) to (t1);
\draw  [->] (t1) to (p3);
\end{tikzpicture}
\caption{A simple net}
\label{net-tau1}
\end{figure}
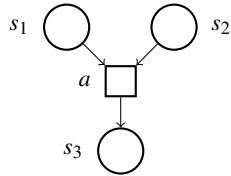

\begin{example}\label{primo-tau-ex}
Consider the simple P/T net in Figure \ref{net-tau1}, with $S = \{s_1, s_2, s_3\}$. It is rather easy to realize that there are only two
interesting (i.e., useful to relate non-deadlocked markings) maximal place bisimulations, namely:

$\begin{array}{lclll}
R_1 & = & \mathcal{I}_S = \{(s_1, s_1), (s_2, s_2), (s_3, s_3)\} \mbox{ and}\\
R_2 & = &  (R_1 \setminus  \mathcal{I}_{\{s_1, s_2\}}) \cup \{(s_1, s_2), (s_2, s_1)\} = \{(s_1, s_2), (s_2, s_1), (s_3, s_3)\},\\
\end{array}$

\noindent
only one of which is an equivalence relation. However, note that 
their union $R = R_1 \cup R_2$ is not a place bisimulation. In fact, on the one hand $(s_1 \oplus s_1, s_1 \oplus s_2) \in R^\oplus$,
but, on the other hand, these two markings do not satisfy the place bisimulation game, because 
$s_1 \oplus s_1$ is stuck, while $s_1 \oplus s_2$ can fire 
the $a$-labeled transition, reaching $s_3$.

As $R_1$ and $R_2$ are two maximal place bisimulations, then we can prove that $m \sim_p m$ for each marking 
$m \in \mathcal{M}(S)$
because $(m, m) \in R_1^\oplus$, as well as that, e.g., $2 \cdot s_1 \oplus s_2 \sim_p s_1 \oplus 2 \cdot s_2$ as
$(2 \cdot s_1 \oplus s_2, s_1 \oplus 2 \cdot s_2) \in R_2^\oplus$.

There are also other four maximal place bisimulations, which however may be used to relate deadlocked markings only:

$\begin{array}{lclll}
R_3 & = & \{(s_1, s_3), (s_3, s_2), (s_3, s_3), (s_1, s_2)\}\\
R_4 & = & \{(s_3, s_1), (s_2, s_3), (s_3, s_3), (s_2, s_1)\}\\
R_5 & = & \{(s_1, s_1), (s_3, s_3), (s_3, s_1), (s_1, s_3)\} \mbox{ and}\\
R_6 & = & \{(s_2, s_2), (s_3, s_3), (s_3, s_2), (s_2, s_3)\},\\
\end{array}$

\noindent
so that, e.g., $(2 \cdot s_1 \oplus s_3, s_2 \oplus 2 \cdot s_3) \in R_3^\oplus$.
Note that the equivalence relation $\sim_{p_{eq}}$ characterized in \cite{AS92} is $R_1^\oplus$, which is an equivalence relation indeed,
because $R_1$ is an equivalence and, by Proposition \ref{add-prop1}(1), also $R_1^\oplus$ is an equivalence.
Instead 
$\sim_p = R_1^\oplus \cup R_2^\oplus \cup R_3^\oplus \cup R_4^\oplus \cup R_5^\oplus \cup R_6^\oplus$, 
so that these two equivalences are very different. 
Note that $\sim_p$ is an equivalence relation, indeed, as
it is reflexive ($\mathcal{I}_M =  (\mathcal{I}_S)^\oplus = R_1^\oplus \subseteq \sim_p$), symmetric (because 
the inverse relation of a maximal place 
bisimulation is a maximal place bisimulation; e.g., $R_3^{-1} = R_4$) and
transitive (maximal place relations are closed by relational composition; e.g., $R_2 \circ R_2 = R_1$ and $R_6 \circ R_5 = R_4$).
\fine
\end{example}

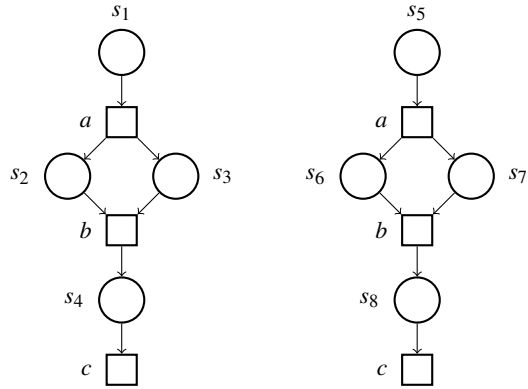
\begin{figure}[t]
\centering
\begin{tikzpicture}[
every place/.style={draw,thick,inner sep=0pt,minimum size=6mm},
every transition/.style={draw,thick,inner sep=0pt,minimum size=4mm},
bend angle=30,
pre/.style={<-,shorten <=1pt,>=stealth,semithick},
post/.style={->,shorten >=1pt,>=stealth,semithick}
]
\def\eofigdist{3.3cm}
\def\eodist{0.4cm}
\def\eodisty{0.8cm}


\node (p1) [place]  [label=above:$s_1$] {};
\node (t1) [transition] [below =\eodist of p1,label=left:$a\;$] {};
\node (p2) [place] [below left =\eodist of t1,label=left:$s_2\;$] {};
\node (p3) [place] [below right =\eodist of t1,label=right:$\;s_3$] {};
\node (t2) [transition] [below right=\eodist of p2,label=left:$b\;$] {};
\node (p4) [place] [below =\eodist of t2,label=left:$s_4\;$] {};
\node (t3) [transition] [below =\eodist of p4,label=left:$c\;$] {};

\draw  [->] (p1) to (t1);
\draw  [->] (t1) to (p2);
\draw  [->] (t1) to (p3);
\draw  [->] (p2) to (t2);
\draw  [->] (p3) to (t2);
\draw  [->] (t2) to (p4);
\draw  [->] (p4) to (t3);


\node (p5) [place]  [right=\eofigdist of p1, label=above:$s_5$] {};
\node (t4) [transition] [below =\eodist of p5,label=left:$a\;$] {};
\node (p6) [place]  [below left=\eodist of t4, label=left:$s_6\;$] {};
\node (p7) [place]  [below right =\eodist of t4,label=right:$\;s_7$] {};
\node (t5) [transition] [below left=\eodist of p7,label=left:$b\;$] {};
\node (p8) [place]  [below =\eodist of t5,label=left:$s_8\;$] {};
\node (t6) [transition] [below=\eodist of p8,label=left:$c\;$] {};

\draw  [->] (p5) to (t4);
\draw  [->] (t4) to (p6);
\draw  [->] (t4) to (p7);
\draw  [->] (p6) to (t5);
\draw  [->] (p7) to (t5);
\draw  [->] (t5) to (p8);
\draw  [->] (p8) to (t6);

\end{tikzpicture}
\caption{A more complex net}
\label{net-dec-place}
\end{figure}

\begin{example}\label{ex-dec}
Consider the net in Figure \ref{net-dec-place}, where $S = \{s_1, s_2,$ $ s_3, s_4,$ $ s_5, s_6, s_7, s_8\}$. 
It is not too difficult to realize that the six interesting maximal place bisimulations
are the following:

$\begin{array}{lclll}
R_1 & = & \mathcal{I}_S \cup \{(s_4, s_8), (s_8, s_4)\}, \mbox{ which is an equivalence relation,}\\
R_2 & = & (R_1 \setminus \mathcal{I}_{\{s_2, s_3\}}) \cup \{(s_2, s_3), (s_3, s_2)\},\\
R_3 & = & (R_1 \setminus \mathcal{I}_{\{s_6, s_7\}}) \cup \{(s_6, s_7), (s_7, s_6)\},\\
\end{array}$

$\begin{array}{lclll}
R_4 & = &  (R_1 \setminus \mathcal{I}_{\{s_2, s_3, s_6, s_7\}}) \cup  \{(s_2, s_3), (s_3, s_2), (s_6, s_7), (s_7, s_6)\},\\
\end{array}$

$\begin{array}{lclll}
R_5 & = & (R_1 \setminus \mathcal{I}_{\{s_1, s_2, s_3, s_5, s_6, s_7\}}) \cup \{(s_2, s_6), (s_3, s_7),  (s_6, s_2), (s_7, s_3)\} 
\cup \{(s_1, s_5), (s_5, s_1)\},\\
R_6 & = & (R_1 \setminus \mathcal{I}_{\{s_1, s_2, s_3, s_5, s_6, s_7\}}) \cup \{(s_2, s_7), (s_3, s_6), (s_6, s_3), (s_7, s_2)\}  \cup \{(s_1, s_5), (s_5, s_1)\}.\\[.3cm]
\end{array}$

However, there are also other, less interesting, maximal place bisimulations, which may be used to relate deadlocked markings.
For instance, $R_7 = \{(s_2, s_3), (s_2, s_6), $ $
(s_6, s_7),$ $(s_6, s_2),$ $ (s_4, s_8), (s_8, s_4)\}$, so that $(2 \cdot s_2, s_3 \oplus s_6) \in R_7^\oplus$.
The equivalence relation $\sim_{p_{eq}}$ characterized in \cite{AS92} is $R_1^\oplus$, 
while $\sim_p \supseteq R_1^\oplus \cup R_2^\oplus $ $
\cup R_3^\oplus \cup R_4^\oplus \cup R_5^\oplus \cup R_6^\oplus \cup R_7^\oplus$.
\fine
\end{example}

\begin{figure}[t]
\centering

\begin{tikzpicture}[
every place/.style={draw,thick,inner sep=0pt,minimum size=6mm},
every transition/.style={draw,thick,inner sep=0pt,minimum size=4mm},
bend angle=42,
pre/.style={<-,shorten <=1pt,>=stealth,semithick},
post/.style={->,shorten >=1pt,>=stealth,semithick}
]
\def\eofigdist{5cm}
\def\eodist{0.4cm}
\def\eodisty{1.5cm}

\node (p1) [place]  [label=above:$P_1$] {};
\node (p2) [place]  [right=\eodisty of p1,label=above:$C_1$] {};
\node (t1) [transition] [below=\eodist of p1,label=left:{\em prod}] {};
\node (p3) [place] [below left=\eodist of t1,label=left:$P_1'$] {};
\node (p4) [place] [below right=\eodist of t1,label=right:$D_1$] {};
\node (t2) [transition] [below=\eodist of p3,label=left:{\em prod}] {};
\node (t3) [transition] [below right=\eodist of p4,label=right:{\em del}] {};
\node (p5) [place] [below =\eodist of t3,label=left:$C_1'$] {};
\node (t4) [transition] [right=\eodist of p5,label=right:{\em cons}] {};

\draw  [->] (p1) to (t1);
\draw  [->] (t1) to (p3);
\draw  [->] (t1) to (p4);
\draw  [->] (p3) to (t2);
\draw  [->, bend left] (t2) to (p3);
\draw  [->, bend right] (t2) to (p4);
\draw  [->] (p4) to (t3);
\draw  [->, bend left] (p2) to (t3);
\draw  [->] (t3) to (p5);
\draw  [->] (p5) to (t4);
\draw  [->, bend right] (t4) to (p2);

\node (q1) [place]  [right=\eofigdist of p1,label=above:$P_2$] {};
\node (q2) [place]  [right=\eodisty of q1,label=above:$C_2$] {};
\node (s1) [transition] [below left=\eodist of q1,label=left:{\em prod}] {};
\node (s'1) [transition] [below right=\eodist of q1,label=right:{\em prod}] {};

\node (q3) [place] [below=\eodist of s1,label=left:$D_2'$] {};
\node (q'3) [place] [below=\eodist of s'1,label=left:$D_2''$] {};

\node (s2) [transition] [below=\eodist of q'3,label=left:{\em del}] {};
\node (s'2) [transition] [below right={1cm} of q'3,label=left:{\em del}] {};

\node (q4) [place] [below right={0.8cm} of s2,label=left:$C_2'$] {};
\node (s3) [transition] [right=\eodist of q4,label=below:{\em cons}] {};

\draw  [->, bend left] (q1) to (s1);
\draw  [->, bend right] (q1) to (s'1);
\draw  [->, bend left] (s1) to (q1);
\draw  [->, bend right] (s'1) to (q1);
\draw  [->] (s1) to (q3);
\draw  [->] (s'1) to (q'3);
\draw  [->] (q3) to (s2);
\draw  [->] (q2) to (s2);
\draw  [->] (q'3) to (s'2);
\draw  [->] (q2) to (s'2);
\draw  [->] (s2) to (q4);
\draw  [->] (s'2) to (q4);
\draw  [->] (q4) to (s3);
\draw  [->, bend right] (s3) to (q2);

\end{tikzpicture}
\caption{Two unbounded producer-consumer systems}
\label{upc1-place}
\end{figure}
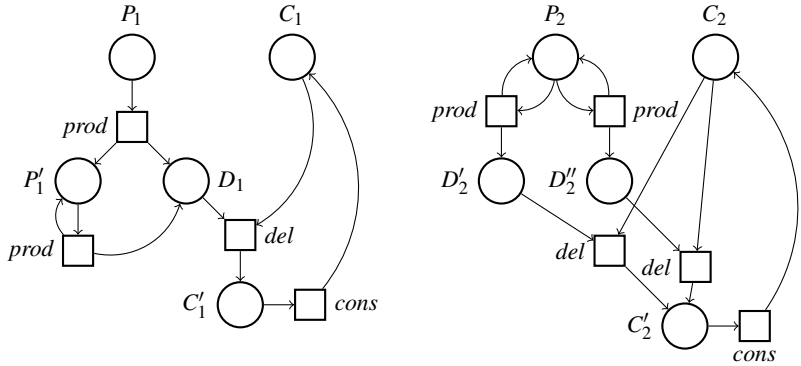

\begin{example}\label{ex-pc}
As a tiny case study, let us consider the two unbounded producer-consumer systems in 
Figure \ref{upc1-place}, where {\em prod} is 
the action of producing an item,
{\em del} of delivering an item, {\em cons} of consuming an item. 
Note that if a token is present in the producer place $P_1'$ (or $P_2$), 
then the transition labeled by $prod$ can be performed at will, with the effect that the number of tokens accumulated into place $D_1$
(or $D_2'$, $D_2''$) can grow unboundedly; hence, the reachable markings from $P_1 \oplus C_1$ (or $P_2 \oplus C_2$)
are infinitely many, so that the state space of these two systems is infinite.
In order to prove that $P_1 \oplus C_1 \sim_p P_2 \oplus C_2$, we can try to build a place bisimulation $R$ which must contain at least the pairs
$(P_1, P_2), (C_1, C_2)$; then,
in order to match the $prod$-labeled transitions, we see that also the pairs $(P_1', P_2),$ $ (D_1, D_2'), $ $(D_1, D_2'')$ are needed, and then,
in order to match the $del$-labeled transitions, also the pair $(C_1', C_2')$ must be considered. The resulting relation is 

$R = \{(P_1, P_2), (P_1', P_2), (D_1, D_2'), (D_1, D_2''), (C_1, C_2), (C_1', C_2')\}$, 

\noindent which is a place bisimulation, indeed 
(formally provable as explained in the subsequent Lemma \ref{pl-rel-dec-lem}), justifying that $P_1 \oplus C_1 \sim_p P_2 \oplus C_2$. 
From $R$ we can also derive that, e.g.,
$P_1 \oplus P_1' \oplus 2 \cdot D_1 \oplus C_1'$ is place bisimilar to $2 \cdot P_2 \oplus D_2' \oplus D_2'' \oplus C_2'$. 
\fine
\end{example}

\begin{figure}[t]
\centering
\begin{tikzpicture}[
every place/.style={draw,thick,inner sep=0pt,minimum size=6mm},
every transition/.style={draw,thick,inner sep=0pt,minimum size=4mm},
bend angle=45,
pre/.style={<-,shorten <=1pt,>=stealth,semithick},
post/.style={->,shorten >=1pt,>=stealth,semithick}
]
\def\eofigdist{5.9cm}
\def\eodist{0.4}
\def\eodisty{1}
\def\eodistw{0.7}

\node (a) [label=left:$a) \quad $]{};

\node (q1) [place] [label={above:$s_1$} ] {};
\node (s1) [transition] [below=\eodist of q1,label=left:$a\;$] {};
\node (q2) [place] [right=\eodisty of q1,label=above:$s_2\;$] {};
\node (s2) [transition] [below left=\eodist of q2,label=right:$\;a$] {};
\node (s3) [transition] [below right=\eodist of q2, label=left:$a\;$] {};
\node (q3) [place] [right=\eodisty of q2,label=above:$\;s_3$] {};
\node (s4) [transition] [below =\eodist of q3,label=right:$\;a$] {};
\node (s5) [transition] [below right=\eodistw of q3,label=right:$\;a$] {};

\node (q4) [place] [below=\eodist of s1,label=right:$\;s_4$] {};
\node (q5) [place] [below=\eodist of s3,label=right:$\;s_5$] {};
\node (s6) [transition] [below =\eodist of q4,label=right:$\;b$] {};
\node (s7) [transition] [below =\eodist of q5,label=right:$\;b$] {};
\node (q6) [place] [below=\eodist of s6,label=below:$\;s_6$] {};
\node (s8) [transition] [left =\eodist of s6,label=left:$b\;$] {};

\node (q7) [place] [right=\eodist of q6,label=below:$\;s_7$] {};
\node (s9) [transition] [right =\eodisty of s6,label=right:$b\;$] {};

\draw  [->] (q1) to (s1);
\draw  [->] (q2) to (s2);
\draw  [->] (q2) to (s3);
\draw  [->] (q3) to (s4);
\draw  [->] (q3) to (s5);
\draw  [->] (s1) to (q4);
\draw  [->] (s2) to (q4);
\draw  [->] (s3) to (q5);
\draw  [->] (s4) to (q5);
\draw  [->, bend left] (s5) to (q7);
\draw  [->] (q4) to (s6);
\draw  [->, bend left] (q5) to (s7);
\draw  [->, bend left] (s7) to node[auto] {2} (q5);
\draw  [->] (s6) to  (q6);
\draw  [->, bend left] (q6) to (s8);
\draw  [->, bend left] (s8) to node[auto] {2} (q4);

\draw  [->] (s6) to  (q7);
\draw  [->, bend left] (q7) to (s9);
\draw  [->, bend left] (s9) to (q7);
\draw  [->] (s9) to  (q4);


\node (b) [right={5.7cm} of a,label=left:$b)\;\;$] {};

\node (p1) [place]  [right=\eofigdist of q1,label=above:$s_8$] {};
\node (t1) [transition] [below=\eodist of p1,label=left:$a\;$] {};
\node (p2) [place]  [below=\eodist of t1,label=right:$s_9$] {};
\node (t2)  [transition] [below=\eodist of p2,label=left:$b\;$] {};

\draw  [->] (p1) to (t1);
\draw  [->] (t1) to (p2);
\draw  [->, bend left] (p2) to (t2);
\draw  [->, bend left] (t2) to node[auto] {2} (p2);

\end{tikzpicture}
\caption{Two place bisimilar nets}
\label{red-fig}
\end{figure}
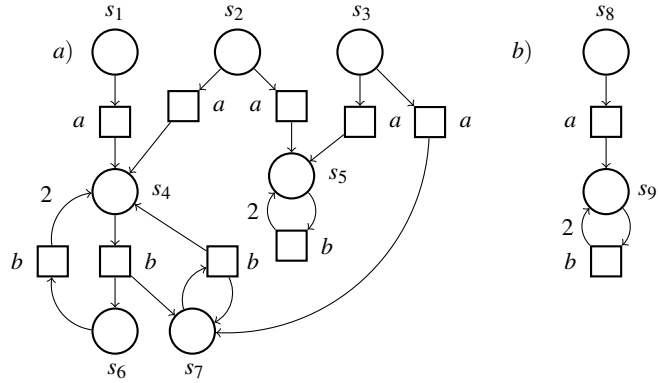

\begin{example}\label{ex-bpp}
As a further example, consider the nets in Figure \ref{red-fig}, that model a simple  system 
with two types of sequential processes: those
that can perform $b$ and, in doing so, generate two instances of the same process type (i.e., $s_4,$ $ s_5, $ $s_6, s_7, s_9$); 
and those
that can perform $a$ and, by doing so, become a process of the first type (i.e., $s_1, s_2, $ $s_3,$$ s_8$).
It is easy to see that $R = \{(s_1,s_8), (s_2,s_8),$ $(s_3,s_8), $  $(s_4,s_9), (s_5,s_9), (s_6,s_9), (s_7,s_9)\}$ is a place bisimulation. Hence, e.g.,
$s_1 \oplus 2 \cdot s_3 \oplus s_4 \oplus s_7$ is place bisimilar to $3 \cdot s_8 \oplus 2 \cdot s_9$. 
\fine
\end{example}

%
\section{Place Bisimilarity Implies Causal-net Bisimilarity} \label{place-cn-bis}
%

Now we want to prove that our definition of place bisimulation does respect the intended causal semantics of Petri nets, 
by proving that place bisimilarity $\sim_p$ implies causal-net bisimilarity $\sim_{cn}$. The reverse implication does not hold, as
illustrated in Example \ref{ex-p>cn}.

\begin{theorem}\label{place-bis>cn-bis}
For each P/T net $N = (S, A, T)$, if $m_1 \sim_p m_2$, then $m_1 \sim_{cn} m_2$.
\proof
If $m_1 \sim_{p} m_2$, then  a place bisimulation $R_1$ exists with
$(m_1, m_2) \in R_1^\oplus$.
Consider 
    \begin{equation*} \label{R2}
        \begin{split}
        R_2 \overset{def}{=} \lbrace (\rho_1, C, \rho_2) | & (C, \rho_1) \text{ is a process of $N(m_{1})$ and} \\
        &(C, \rho_2) \text{ is a process of $N(m_{2})$ and} \\
        & \forall b \in Max(C) \; (\rho_1(b), \rho_2(b)) \in R_1
         \rbrace .
        \end{split}
    \end{equation*}

\noindent
We want to prove that $R_2$ is a causal-net bisimulation.
First of all, consider a triple of the form $(\rho_1^0, C^0, \rho_2^0)$,
where $C^0$ is the causal net without events and $\rho_1^0, \rho_2^0$ are chosen in such a way that, 
not only  $\rho_i^0(Min(C^0)) = \rho_i^0(Max(C^0)) = m_i$ for $i= 1, 2$, but also that 
$(\rho_1^0(b), \rho_2^0(b)) \in R_1$ for all $b \in Max(C^0)$, which is really possible because we know that $(m_1, m_2) \in R_1^\oplus$.
Then $(\rho_1^0, C^0, \rho_2^0)$ must belong to $R_2$,
because $(C^0, \rho_i^0)$ is a process of $N(m_i)$, for $i=1, 2$ and, 
by construction, $(\rho_1^0(b), \rho_2^0(b)) \in R_1$ for all $b \in Max(C^0)$. 
Hence, if $R_2$ is a causal-net bisimulation, then the triple 
$(\rho_1^0, C^0, \rho_2^0) \in R_2$ ensures that $m_1 \sim_{cn} m_2$. 
Now, assume $(\rho_1, C, \rho_2) \in R_2$. 
In order for $R_2$ to be a causal-net bisimulation,
we must prove that
\begin{enumerate}
\item
$\forall t_1, C', \rho_1'$ s.t. $(C, \rho_1) \deriv{e} (C', \rho_1')$,
where $\rho_1'(e) = t_1$,
$\exists t_2, \rho_2'$ s.t.
$(C, \rho_2) \deriv{e} (C', \rho_2')$,
where $\rho_2'(e) = t_2$, and
$(\rho'_1, C', \rho'_2) \in R_2$;
\item symmetrical, if $(C, \rho_2)$ moves first.
\end{enumerate}

Assume $(C, \rho_1) \deriv{e} (C', \rho_1')$ with $\rho_1'(e) = t_1$.
Recall that if $(\rho_1, C, \rho_2)  \in R_2$ then  for all $b \in Max(C)$ we have $(\rho_1(b),$ $ \rho_2(b)) \in R_1$.
This means that $(\rho_1(\pre{e}), \rho_2(\pre{e})) \in R_1^\oplus$. As $R_1$ is a place bisimulation,
to the move $\rho_1(\pre{e})[t_1\rangle \post{t_1} $, marking  $\rho_2(\pre{e})$ can respond
with $\rho_2(\pre{e})[t_2\rangle \post{t_2}$ for some suitable transition $t_2$
such that $(\pre{t_1}, \pre{t_2}) \in R_1^\oplus$, $l(t_1) = l(t_2)$ and $(\post{t_1}, \post{t_2}) \in R_1^\oplus$.
Therefore, since $t_1$ and $t_2$ have the samem shape,
it is possible to derive $(C, \rho_2) \deriv{e} (C', \rho_2')$,
where $\rho_2'$ is chosen in such a way that, not only $\rho_2'(e) = t_2$, but also that $(\rho_1'(b), \rho_2'(b)) \in R_1$ for each $b \in \post{e}$ 
(which is really possible because $(\post{t_1}, \post{t_2}) \in R_1^\oplus$).
Since for all $b \in (Max(C) \ominus \pre{e})$ we already know that  $(\rho_1(b),$ $ \rho_2(b)) \in R_1$,
it follows that
for all $b' \in Max(C')$ it holds that $(\rho_1'(b'), \rho_2'(b')) \in R_1$.
As a consequence, we also have that $(\rho'_1, C', \rho'_2) \in R_2$.

The case when $(C, \rho_2)$ moves first is symmetrical and therefore omitted.
Thus, $R_2$ is a causal-net bisimulation and, since $(\rho_1^0, C^0, \rho_2^0) \in R_2$, we have
$m_1 \sim_{cn} m_2$.
\fine
\end{theorem}

%
\section{Place Bisimilarity is Decidable} \label{decid-place-sec}
%
 
Since the union of place bisimulations may be not a place bisimulation, its definition is not coinductive
and, therefore, the well-known polynomial algorithms \cite{KS83,PT87} for computing the largest bisimulation on LTSs 
cannot be adapted to compute place bisimilarity.

In order to prove that $\sim_p$ is decidable, we first need a technical lemma which states that it is decidable to check 
whether a place relation $R \subseteq S \times S$ is a place bisimulation.

\begin{lemma}\label{pl-rel-dec-lem}
Given a P/T net $N = (S, A, T)$ and a place relation $R \subseteq S \times S$, it is decidable if $R$ 
is a place bisimulation.
\proof
We want to prove that $R$ is a place bisimulation if and only if the following two conditions are satisfied:
\begin{enumerate}
\item $\forall t_1 \in T$, $\forall m$ such that $(\pre{t_1}, m) \in R^\oplus$
	\begin{itemize}
	\item[$(a)$] $\exists t_2$ such that $m = \pre{t_2}$, with  $l(t_1) = l(t_2)$ and $(\post{t_1}, \post{t_2}) \in R^\oplus$. 
	\end{itemize}
\item $\forall t_2 \in T$, $\forall m$ such that $(m, \pre{t_2}) \in R^\oplus$
	\begin{itemize}
	\item[$(b)$] $\exists t_1$ such that $m = \pre{t_1}$,
	with $l(t_1) = l(t_2)$ and $(\post{t_1}, \post{t_2}) \in R^\oplus$.
	\end{itemize}
\end{enumerate}

The implication from left to right is obvious: if $R$ is a place bisimulation, then for sure conditions 1 and 2 are satisfied.
For the converse implication, assume that conditions 1 and 2 are satisfied; then we have to prove that the 
place bisimulation game for $R$ holds for all pairs $(m_1, m_2) \in R^\oplus$.

Let $ q = \{(s_1, s_1'), (s_2, s_2'), \ldots,$ $(s_k, s_k')\}$ be any multiset of associations
that can be used to prove that $(m_1, m_2) \in R^\oplus$. So this means that 
$m_1 = s_1 \oplus s_2 \oplus \ldots \oplus s_k$, $m_2 = s_1' \oplus s_2' \oplus \ldots \oplus s_k'$
and that $(s_i, s_i') \in R$ for $i = 1, \ldots, k$. 

If $m_1 [t_1 \rangle m_1'$, then $m_1' = m_1 \ominus \pre{t_1} \oplus \post{t_1}$.
Consider any multiset of associations $p = \{(\overline{s}_{1}, \overline{s}'_{1}),$ $\ldots, (\overline{s}_{h}, \overline{s}'_{h})\} \subseteq q$,
with $\overline{s}_{1} \oplus  \ldots \oplus \overline{s}_{h} $ $= \pre{t_1}$.
Note that $(\pre{t_1}, \overline{s}'_{1} \oplus  \ldots \oplus \overline{s}'_{h}) \in R^\oplus$. 
Therefore, by condition 1, there exists a transition $t_{2}$ such that
$\pre{t_{2}} = \overline{s}'_{1} \oplus  \ldots \oplus \overline{s}'_{h}$, $l(t_1) = l(t_{2})$ and $(\post{t_1}, \post{t_{2}}) \in R^\oplus$.
Hence, since $\pre{t_2} \subseteq m_2$, also $m_2 [t_{2} \rangle m_2'$ is firable, where $m_2' = m_2 \ominus \pre{t_{2}} \oplus \post{t_{2}}$, so that
$(\pre{t_1}, \pre{t_{2}}) \in R^\oplus$,  $(\post{t_1}, \post{t_{2}}) \in R^\oplus$, $l(t_1) = l(t_{2})$ and, 
finally, $(m_1', m_2') \in R^\oplus$, as required, which holds because, from the set $q$
of matching pairs for $m_1$ and $m_2$,
we have removed those in $p$ and we have added
those justifying $(\post{t_1}, \post{t_{2}}) \in R^\oplus$. 

If $m_2 [t_2 \rangle m_2'$, then we have to use an argument symmetric to the above, where condition 2 is used instead.
Hence, we have proved that conditions 1 and 2 are enough to prove that $R$ is a place bisimulation.

Finally, the complexity of this procedure is as follows. For condition 1, 
we have to consider all the net transitions,
and for each $t_1$ we have to consider
all the markings $m$ that $(\pre{t_1}, m) \in R_i^\oplus$, and for each pair $(t_1, m)$
we have to check whether there exists a transition $t_2$ such that $m = \pre{t_2}$, $l(t_1) = l(t_2)$ and
$(\post{t_1}, \post{t_2}) \in R_i^\oplus$. And the same for condition 2. Hence, this  
procedure has worst-case time complexity 
$O(q \cdot n^{p_1} \cdot q  \cdot  (p_2^2\sqrt{p_2}))$,
where $q = |T|$, $n = |S|$ and  $p_1, p_2$ are the least numbers such that $| \pre{t}| \leq p_1$ 
and $|\post{t}| \leq p_2$ for all $t \in T$, as the number of markings $m$ related via $R_i$ to $\pre{t_1}$ is  $n^{p_1}$ at most, 
and checking whether $(\post{t_1}, \post{t_2}) \in R_i^\oplus$ takes
$O(p_2^2\sqrt{p_2})$ in the worst-case.
\fine
\end{lemma}

\begin{theorem}\label{pl-bis-decid-th}{\bf (Place bisimilarity is decidable)}
Given a P/T net $N = (S, A, T)$, for each pair of markings $m_1$ and $m_2$, it is decidable whether $m_1 \sim_p m_2$.
\proof
If $|m_1| \neq |m_2|$, then $m_1 \nsim_p m_2$ by Proposition \ref{fin-k-add}. Otherwise, we can assume that $|m_1| = k = |m_2|$.
Since $S$ is finite, the set of all the place relations over $S$ is finite as well. Let us list all the place relations as follows: $R_1, R_2, \ldots, R_n$.
Therefore, for $i = 1, \ldots, n$, by Lemma \ref{pl-rel-dec-lem} we can decide whether the place relation $R_i$ is 
a place bisimulation and, in such a case,
we can check whether $(m_1, m_2) \in R_i^\oplus$ in $O(k^2\sqrt{k})$ time. 
As soon as we have found a place bisimulation $R_i$ such that $(m_1, m_2) \in R_i^\oplus$,
we stop concluding that $m_1 \sim_p m_2$. If none of the $R_i$ is a place bisimulation such that $(m_1, m_2) \in R_i^\oplus$, then
we can conclude that $m_1 \nsim_p m_2$. 
\fine
\end{theorem}

About the complexity of this decision procedure, we note that if $|S| = n$, then the set of all the place relations
over $S$ has cardinality $2^{n^2}$. Moreover, the procedure for checking whether a place relation $R_i$ in this 
set is a place bisimulation 
has worst-case time complexity $O(q \cdot n^{p_1} \cdot q  \cdot  (p_2^2\sqrt{p_2}))$.
Finally, if $R_i$ is a place bisimulation, the cost of checking whether $(m_1, m_2) \in R_i^\oplus$ is
$O(k^2 \sqrt{k})$ if the two markings have size $k$. Summing up, 
the procedure is exponential in the size of the net.

\section{A Coarser Variant: D-place Bisimilarity} \label{d-place-sec} 

Now we want to define a slightly weaker variant of place bisimulation, we call {\em d-place bisimulation},
which may relate a place $s$ also to the empty marking $\theta$.
In order to provide the definition of d-place bisimulation, 
we need first to extend the domain of a place relation: 
the empty marking $\theta$ is considered as an additional place, so that a place 
relation is defined not on $S$, rather on $S \cup \{\theta\}$.
Hence, 
the symbols $r_1$ and $r_2$ that occur in the following definitions, 
can only denote either the empty marking $\theta$ or a 
single place $s$. 
Now we extend the idea of additive closure to these more general place relations, 
yielding {\em d-additive closure}.

\begin{definition}\label{hadd-eq}{\bf (D-additive closure)}
Given a P/T net $N = (S, A, T)$ and a {\em place relation} $R \subseteq (S\cup \{\theta\}) \times (S \cup \{\theta\})$, we define 
a {\em marking relation}
$R^\odot \, \subseteq \, {\mathcal M}(S) \times {\mathcal M}(S)$, called 
the {\em d-additive closure} of $R$,
as the least relation induced by the following axiom and rule.

$\begin{array}{lllllllllll}
 \bigfrac{}{(\theta, \theta) \in  R^\odot} & \; & \; 
 \bigfrac{(r_1, r_2) \in R \; \; (m_1, m_2) \in R^\odot }{(r_1 \oplus m_1, r_2 \oplus m_2) \in  R^\odot }  \\
\end{array}$
\\[-.2cm]
\fine
\end{definition}

Note that if two markings are related by $R^\odot$, then they 
may have different size; 
in fact, even if the axiom relates the empty marking to itself (so two markings with the same size),
as $R \subseteq (S\cup \{\theta\}) \times (S \cup \{\theta\})$, it may be the case that $(\theta, s) \in R$,
so that, assuming $(m_1', m_2') \in R^\odot$ with $|m_1'| = |m_2'|$, we get that the pair
$(m_1', s \oplus m_2')$ belongs to $R^\odot$, as $\theta$ is the identity for
the operator of multiset union. 
Hence, Proposition \ref{fin-k-add}, which is valid for place relations defined over $S$, is not valid for place relations 
defined over $S\cup \{\theta\}$. 
However, the properties in Propositions \ref{add-prop1}
and \ref{add-prop2} hold also for these more general place relations.
Note that checking whether $(m_1, m_2) \in R^\odot$ has complexity $O(k^2\sqrt{k})$, where $k$ is the size of the largest marking.

\begin{definition}\label{def-dplace-bis}{\bf (D-place bisimulation)}
Let $N = (S, A, T)$ be a P/T net. 
A {\em d-place bisimulation} is a relation
$R\subseteq (S\cup \{\theta\}) \times (S \cup \{\theta\})$ such that if $(m_1, m_2) \in R^\odot$
then
\begin{itemize}
\item $\forall t_1$ such that  $m_1[t_1\rangle m'_1$, $\exists t_2$ such that $m_2[t_2\rangle m'_2$ 
with $(\pre{t_1}, \pre{t_2}) \in R^\odot$, $l(t_1) = l(t_2)$,  $(\post{t_1}, \post{t_2}) \in R^\odot$ and, moreover, 
$(m_1', m_2') \in R^\odot$,
\item $\forall t_2$ such that  $m_2[t_2\rangle m'_2$, $\exists t_1$ such that $m_1[t_1\rangle m'_1$ 
with $(\pre{t_1}, \pre{t_2}) \in R^\odot$, $l(t_1) = l(t_2)$,  $(\post{t_1}, \post{t_2}) \in R^\odot$ and, moreover, 
$(m_1', m_2') \in R^\odot$.
\end{itemize}

Two markings $m_1$ and $m_2$ are  {\em d-place bisimilar}, denoted by
$m_1 \sim_{d} m_2$, if there exists a d-place bisimulation $R$ such that $(m_1, m_2) \in R^\odot$.
\fine
\end{definition}

\begin{figure}[!t]
\centering
\begin{tikzpicture}[
every place/.style={draw,thick,inner sep=0pt,minimum size=6mm},
every transition/.style={draw,thick,inner sep=0pt,minimum size=4mm},
bend angle=30,
pre/.style={<-,shorten <=1pt,>=stealth,semithick},
post/.style={->,shorten >=1pt,>=stealth,semithick}
]
\def\eofigdist{3.2cm}
\def\eodist{0.3cm}
\def\eodisty{0.8cm}

\node (a) [label=left:$a)\qquad $]{};

\node (p1) [place]  [label=above:$s_1$] {};
\node (t1) [transition] [below =\eodist of p1,label=left:$a\;$] {};
\node (p2) [place] [below =\eodist of t1,label=left:$s_2\;$] {};
\node (t2) [transition] [below =\eodist of p2,label=left:$b\;$] {};
\node (p3) [place] [below  =\eodist of t2,label=left:$s_3\;$] {};

\draw  [->] (p1) to (t1);
\draw  [->] (t1) to (p2);
\draw  [->] (p2) to (t2);
\draw  [->] (t2) to (p3);

  \node (b) [right={3.1cm} of a,label=left:$b)\quad$] {};

\node (p4) [place]  [right=\eofigdist of p1, label=above:$s_4$] {};
\node (t3) [transition] [below  =\eodist of p4,label=left:$a\;$] {};
\node (p5) [place]  [below left =\eodist of t3, label=left:$s_5\;$] {};
\node (p6) [place]  [below right =\eodist of t3, label=right:$\;s_6$] {};
\node (t4) [transition] [below =\eodist of p6,label=left:$b\;$] {};

\draw  [->] (p4) to (t3);
\draw  [->] (t3) to (p5);
\draw  [->] (t3) to (p6);
\draw  [->] (p6) to (t4);

\end{tikzpicture}
\caption{Two d-place bisimilar nets}
\label{net-d-place}
\end{figure}
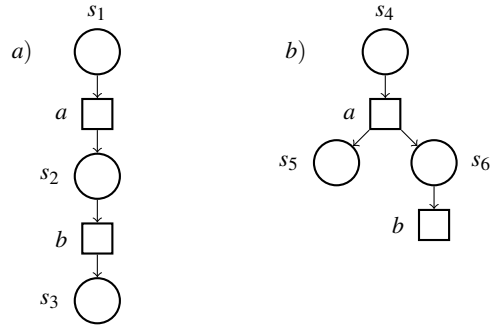

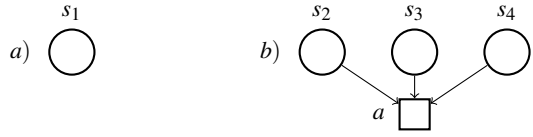
\begin{figure}[!t]
\centering
\begin{tikzpicture}[
every place/.style={draw,thick,inner sep=0pt,minimum size=6mm},
every transition/.style={draw,thick,inner sep=0pt,minimum size=4mm},
bend angle=30,
pre/.style={<-,shorten <=1pt,>=stealth,semithick},
post/.style={->,shorten >=1pt,>=stealth,semithick}
]
\def\eofigdist{2.7cm}
\def\eodist{0.3cm}
\def\eodisty{0.6cm}

\node (a) [label=left:$a)\quad $]{};

\node (p1) [place]  [label=above:$s_1$] {};

  \node (b) [right={3.1cm} of a,label=left:$b)\quad$] {};

\node (p2) [place]  [right=\eofigdist of p1, label=above:$s_2$] {};
\node (p3) [place]  [right=\eodisty of p2, label=above:$s_3$] {};
\node (p4) [place]  [right=\eodisty of p3, label=above:$s_4$] {};
\node (t2) [transition] [below =\eodist of p3,label=left:$a\;$] {};

\draw  [->] (p2) to (t2);
\draw  [->] (p3) to (t2);
\draw  [->] (p4) to (t2);

\end{tikzpicture}
\caption{Relation $\{(s_1, s_2), (\theta, s_3)\}$ is a d-place bisimulation}
\label{net-d3-place}
\end{figure}

\begin{example}\label{ex-dplace}
Consider the net in Figure \ref{net-d-place}. It is easy to realize that $R = \{(s_1, s_4), (\theta, s_5),$ $
(s_2, s_6),$ $(s_3, \theta)\}$ is a d-place bisimulation. Hence, this example shows that d-place bisimilarity is strictly coarser
than place bisimilarity, and not finer than
causal-net bisimilarity, because $s_1$ and $s_4$ generate different causal nets.

The places that are related to $\theta$ (i.e., $s_3$ and $s_5$) are deadlocks, i.e., they have empty post-set.
However, it may happen that a d-place bisimulation can also relate a place with non-empty post-set to $\theta$.
In fact, consider the net in Figure \ref{net-d3-place}. It is easy to observe that the relation 
$R = \{(s_1, s_2), (\theta, s_3)\}$ is a d-place bisimulation, as for all the pairs $(m_1, m_2) \in R^\odot$,
both markings are stuck, so that the d-place bisimulation game is vacuously satisfied.

Nonetheless, it is easy to observe that
if a d-place bisimulation $R$ relates a place $s$ with non-empty post-set to $\theta$, then it is not possible
to find two transitions $t_1$ and $t_2$ such that for the proof of $(\pre{t_1}, \pre{t_2}) \in R^\odot$ it is 
necessary to use the pair $(s, \theta)$ (cf. the subsequent Example \ref{ex-dead}). In other words,
the condition $(\pre{t_1}, \pre{t_2}) \in R^\odot$ in Definition \ref{def-dplace-bis} is actually $(\pre{t_1}, \pre{t_2}) \in R^\oplus$.
\fine
\end{example}

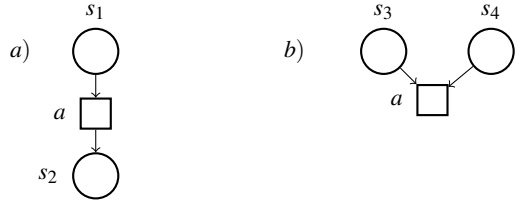
\begin{figure}[!t]
\centering
\begin{tikzpicture}[
every place/.style={draw,thick,inner sep=0pt,minimum size=6mm},
every transition/.style={draw,thick,inner sep=0pt,minimum size=4mm},
bend angle=30,
pre/.style={<-,shorten <=1pt,>=stealth,semithick},
post/.style={->,shorten >=1pt,>=stealth,semithick}
]
\def\eofigdist{3.2cm}
\def\eodist{0.3cm}
\def\eodisty{0.8cm}

\node (a) [label=left:$a)\qquad $]{};

\node (p1) [place]  [label=above:$s_1$] {};
\node (t1) [transition] [below =\eodist of p1,label=left:$a\;$] {};
\node (p2) [place] [below =\eodist of t1,label=left:$s_2\;$] {};

\draw  [->] (p1) to (t1);
\draw  [->] (t1) to (p2);

  \node (b) [right={3.1cm} of a,label=left:$b)\quad$] {};

\node (p3) [place]  [right=\eofigdist of p1, label=above:$s_3$] {};
\node (p4) [place]  [right=\eodisty of p3, label=above:$s_4$] {};
\node (t2) [transition] [below right =\eodist of p3,label=left:$a\;$] {};

\draw  [->] (p3) to (t2);
\draw  [->] (p4) to (t2);

\end{tikzpicture}
\caption{Two fc-bisimilar nets, but not d-place bisimilar}
\label{net-d2-place}
\end{figure}

\begin{example}\label{ex-dead}
Consider Figure \ref{net-d2-place}. Even if $s_1$ and $s_3 \oplus s_4$
are fc-bisimilar, we cannot find any d-place bisimulation relating these two markings. If we include the necessary 
pairs $(s_1, s_3)$ and $(\theta, s_4)$, then we would fail immediately, because the pair $(s_1, s_3)$ does not satisfy
the d-place bisimulation game, as $s_1$ can move, while $s_3$ cannot.
\fine
\end{example}

In order to prove that $\sim_d$ is an equivalence relation, we now list some useful properties of d-place bisimulation relations,
whose proof is omitted because very similar to that of the analogous Proposition \ref{pt-prop-bis}. 

\begin{proposition}\label{pt-d-prop-bis}
For each P/T net $N = (S, A, T)$, the following hold:
\begin{enumerate}
\item The identity relation ${\mathcal I}_S = \{ (r, r) \mid r \in S \cup \{\theta\} \}$ is a d-place bisimulation;
\item the inverse relation $R^{-1} = \{ (r', r) \mid (r, r') \in R\}$ of a d-place bisimulation $R$ is a d-place bisimulation;
\item the relational composition $R_1 \circ R_2 = \{ (r, r'') \mid $ $\exists r'. (r, r') \in R_1 \wedge (r', r'') \in R_2 \}$ of
two d-place bisimulations $R_1$ and $R_2$ is a d-place bisimulation.
\end{enumerate}
\end{proposition}

Hence, in a standard way, we can prove that d-place bisimilarity $\sim_d$ is an equivalence relation, too.

%
\subsection{D-place Bisimilarity Implies Fully-concurrent Bisimilarity} \label{place-fc-bis}
%

D-place bisimilarity $\sim_d$ is incomparable w.r.t. causal-net bisimilarity 
(cf. Example \ref{ex-p>cn} and \ref{ex-dplace}). 
However, we now show that $\sim_d$ does fully respect causality and branching time, 
by proving that $\sim_d$ implies fully-concurrent bisimilarity $\sim_{fc}$. Example \ref{ex-p-fc} and Example \ref{ex-dead} show 
that the implication is strict. 

In the following theorem, given a causal net $C$ and a condition $b \in Max(C)$, with the notation  $\widehat{E_{b}}$ 
we consider the partial order of all the events $e \in E$ such that $e \mathsf{F}^+ b$, where $\mathsf{F}$ is the flow relation for $C$.

\begin{theorem}\label{d-place-bis>fc-bis}
For each P/T net $N = (S, A, T)$, if $m_1 \sim_d m_2$, then $m_1 \sim_{fc} m_2$.
\proof
    If $m_1 \sim_{d} m_2$, then there exists a d-place bisimulation $R_1$ such that
$(m_1, m_2) \in R_1^\odot$.
Let us consider 
    \begin{equation*} \label{R2}
        \begin{split}
        R_2 \overset{def}{=} \lbrace (\pi_1, g, \pi_2) | & \pi_1 = (C_1, \rho_1) \text{ is a process of $N(m_{1})$,} \\
        & \pi_2 = (C_2, \rho_2) \text{ is a process of $N(m_{2})$},\\ 
        & \text{$g$ is an event isomorphism between $\mathsf{E}_{C_1}$ and $\mathsf{E}_{C_2}$},\\ 
         & \text{and property } \Phi(\pi_1, g, \pi_2) \text{ holds}
        \rbrace ,
        \end{split}
    \end{equation*}

\noindent
where property $\Phi(\pi_1, g, \pi_2)$ states that there exists a multiset 

$q = \{(r_1, r_1'),$ $ (r_2, r_2'),$ $ \ldots,$ $(r_k, r_k')\}$

\noindent
of associations such that if $Max(C_1) = b_1 \oplus \ldots \oplus b_{k_1}$ and $Max(C_2) = b_1' 
\oplus \ldots \oplus b'_{k_2}$ (with $k_1, k_2 \leq k$)
then we have that 
\begin{enumerate}
\item $\rho_1(Max(C_1)) = r_1 \oplus \ldots \oplus r_k$ and $\rho_2(Max(C_2)) = r_1' \oplus \ldots \oplus r_k'$ 
(remember that some of the $r_i$ or $r_i'$ can be $\theta$),
\item for $i = 1, \ldots, k$, $(r_i, r_i') \in R_1$, so that $(\rho_1(Max(C_1)), \rho_2(Max(C_2))) \in R_1^\odot$,
\item and for $i = 1, \ldots, k$, if $r_i = \rho_1(b_j)$  for some $b_j \in Max(C_1)$, then either $r_i' = \theta$, 
or $r_i' = \rho_2(b'_{j'})$ for some $b'_{j'} \in Max(C_2)$ such that the partial orders of their predecessors, $\widehat{E^1_{b_j}}$ and
$\widehat{E^2_{b'_{j'}}}$, respectively, are isomorphic via $g$. 
And symmetrically, if $r_i' = \rho_2(b'_{j'})$ for some 
$b'_{j'} \in Max(C_2)$, then
either $r_i = \theta$, or $r_i = \rho_1(b_j)$ for some $b_j \in Max(C_1)$  such that the partial orders of their predecessors, 
$\widehat{E^1_{b_j}}$ and $\widehat{E^2_{b'_{j'}}}$, respectively, are isomorphic via $g$. 
\end{enumerate}

Note that such a multiset $q$ has the property that for each $(r_i, r_i') \in q$, we have that either one of the two elements in the pair is
$\theta$, or both places are the image of suitable minimal conditions, or both places are the image of conditions generated by events related by 
the event isomorphism $g$.

We want to prove that $R_2$ is a fully-concurrent bisimulation.
First of all, consider a triple of the form $(\pi_1^0, g^0, \pi_2^0)$, 
where $\pi_i^0 = (C_1^0, \rho_i^0)$, 
$C_i^0$ is the causal net without events,  $g^0$ is the empty function and $\rho_1^0, \rho_2^0$ are chosen in such a way that, not only 
$\rho_i^0(Min(C_i^0)) = \rho_i^0(Max(C_i^0)) = m_i$ for $i= 1, 2$, but also that, if
$\rho^0_1(Max(C^0_1)) = r_1 \oplus \ldots \oplus r_k$ and $\rho^0_2(Max(C^0_2)) = r_1' \oplus \ldots \oplus r_k'$ 
(remember that some of the $r_i$ or $r_i'$ can be $\theta$), then
$(r_i, r_i') \in R_1$ for $i = 1, \ldots, k$, so that $(\rho^0_1(Max(C^0_1)), \rho^0_2(Max(C^0_2))) \in R_1^\odot$,
which is really possible because, by hypothesis, we have that $(m_1, m_2) \in R_1^\odot$.
Then $(\pi_1^0, g^0, \pi_2^0)$ must belong to $R_2$,
because $(C_i^0, \rho_i^0)$ is a process of $N(m_i)$, for $i=1, 2$ and $\Phi(\pi_1^0, g^0, \pi_2^0)$
trivially holds. 
Hence, if $R_2$ is a fully-concurrent bisimulation, then the triple 
$(\pi_1^0, g^0, \pi_2^0) \in R_2$ ensures that $m_1 \sim_{fc} m_2$. 

Now, assume $(\pi_1, g, \pi_2) \in R_2$, so that property $\Phi(\pi_1, g, \pi_2)$ holds.
In order for $R_2$ to be an fc-bisimulation,
we must prove that
\begin{itemize}
\item[$i)$] 
$\forall t_1, \pi_1'$ such that $\pi_1 \deriv{e_1} \pi_1'$ with $\rho_1'(e_1) = t_1$, $\exists t_2, \pi_2', g'$ such that

\begin{enumerate}
\item $\pi_2 \deriv{e_2} \pi_2'$ with $\rho_2'(e_2) = t_2$;
 \item $g' = g \cup \{(e_1, e_2)\}$, and finally,
\item $(\pi_1', g', \pi_2') \in R_2$;
\end{enumerate}

\item[$ii)$] and symmetrically, if $\pi_2$ moves first.
\end{itemize}
\noindent
Assume $\pi_1 = (C_1, \rho_1) \deriv{e_1} (C_1', \rho_1') = \pi_1'$ with $\rho_1'(e_1) = t_1$.
Now, let  $p = \{(\overline{r}_{1}, \overline{r}'_{1}),$ $\ldots,$ $ (\overline{r}_{h}, \overline{r}'_{h})\} \subseteq q$,
with $\overline{r}_{1}\oplus  \ldots \oplus \overline{r}_{h}$ $= \pre{t_1}$.
Note that $(\pre{t_1}, \overline{r}'_{1} \oplus  \ldots \oplus \overline{r}'_{h}) \in R_1^\odot$. Since $R_1$ is a d-place bisimulation,
there exists $t_2$ such that $\pre{t_2} = \overline{r}'_{1} \oplus  \ldots \oplus \overline{r}'_{h}$, $l(t_1) = l(t_2)$,
$(\post{t_1}, \post{t_2}) \in R_1^\odot$. Therefore, because property $\Phi(\pi_1, g, \pi_2)$ holds, 
it is possible to single out an event $e_2$, with $\rho_2(\pre{e_2}) = \pre{t_2}$, 
such that  $\pi_2 = (C_2, \rho_2) \deriv{e_2} (C_2', \rho_2') = \pi_2'$
(where $\rho_2'$ is such that $\rho_2'(e_2) = t_2$) and such that
the set of events generating the conditions of $\pre{e_1}$ (mapped by $\rho_1$ to $\pre{t_1}$) are isomorphic, via $g$,
to the set of events generating the conditions of $\pre{e_2}$ (mapped by $\rho_2$ to $\pre{t_2}$). Hence, 
the new generated events $e_1$ and $e_2$ have isomorphic predecessors via $g$. So, by defining $g' = g \cup \{(e_1, e_2)\}$,
we conclude that $g'$ is an event isomorphism between $E_{C_1'}$ and $E_{C_2'}$.

Finally, from the multiset of associations $q$ we remove the associations in $p$ and add any multiset $p'$ of associations
that can be used to prove that $(\post{t_1}, \post{t_2}) \in R_1^\odot$. This new multiset $q'$ can be used to prove
that property $\Phi(\pi_1', g', \pi_2')$ holds, because by construction  we have that
$(\rho_1'(\post{e_1}), \rho_2'(\post{g'(e_1)})) \in R_1^\odot$. 
As a consequence we get that $(\pi'_1, g', \pi'_2) \in R_2$.

The case when $\pi_2 = (C_2, \rho_2)$ moves first is symmetrical and so omitted.
Thus, $R_2$ is a fully-concurrent bisimulation and, since $(\pi_1^0, g^0, \pi_2^0) \in R_2$, we have
$m_1 \sim_{fc} m_2$.
\fine
\end{theorem}

\subsection{D-place bisimilarity is decidable}\label{d-place-dec-sec}

Now, we follow the same steps of the decidability proof of $\sim_p$ to prove that also d-place bisimilarity is decidable.
Since the proof is very similar, we simply sketch it.

\begin{lemma}\label{dpl-rel-dec-lem}
Given a P/T net $N = (S, A, T)$ and a relation $R \subseteq (S \cup \{\theta\}) \times (S \cup \{\theta\})$, it is decidable if $R$ 
is a d-place bisimulation.
\proof
We want to prove that $R$ is a d-place bisimulation if and only if the following two conditions are satisfied:
\begin{enumerate}
\item $\forall t_1 \in T$, $\forall m$ such that $(\pre{t_1}, m) \in R^\odot$
	\begin{itemize}
	\item[$(a)$] $\exists t_2$ such that $m = \pre{t_2}$, with  $l(t_1) = l(t_2)$ and $(\post{t_1}, \post{t_2}) \in R^\odot$. 
	\end{itemize}
\item $\forall t_2 \in T$, $\forall m$ such that $(m, \pre{t_2}) \in R^\odot$
	\begin{itemize}
	\item[$(b)$] $\exists t_1$ such that $m = \pre{t_1}$,
	with $l(t_1) = l(t_2)$ and $(\post{t_1}, \post{t_2}) \in R^\odot$.
	\end{itemize}
\end{enumerate}

The implication from left to right is obvious: if $R$ is a d-place bisimulation, then for sure conditions 1 and 2 are satisfied.
For the converse implication, assume that conditions 1 and 2 are satisfied; then we have to prove that the 
d-place bisimulation game for $R$ holds for all pairs $(m_1, m_2) \in R^\odot$. 

Let $ q = \{(r_1, r_1'), (r_2, r_2'), \ldots,$ $(r_k, r_k')\}$ be any multiset of associations
that can be used to prove that $(m_1, m_2) \in R^\odot$. Note that some of the $r_i$ or of the $r'_i$ can be $\theta$.
So this means that 
$m_1 = r_1 \oplus r_2 \oplus \ldots \oplus r_k$, $m_2 = r_1' \oplus r_2' \oplus \ldots \oplus r_k'$
and that $(r_i, r_i') \in R$ for $i = 1, \ldots, k$. 

If $m_1 [t_1 \rangle m_1'$, then $m_1' = m_1 \ominus \pre{t_1} \oplus \post{t_1}$.
Consider any multiset of associations $p = \{(\overline{r}_{1}, \overline{r}'_{1}),$ $\ldots, (\overline{r}_{h}, \overline{r}'_{h})\} \subseteq q$,
with $\overline{r}_{1}\oplus  \ldots \oplus \overline{r}_{h}$ $= \pre{t_1}$.
Note that $(\pre{t_1}, \overline{r}'_{1} \oplus  \ldots \oplus \overline{r}'_{h}) \in R^\odot$. 
Therefore, by condition 1, there exists a transition $t_{2}$ such that
$\pre{t_{2}} = \overline{r}'_{1} \oplus  \ldots \oplus \overline{r}'_{h}$, $l(t_1) = l(t_{2})$ and $(\post{t_1}, \post{t_{2}}) \in R^\odot$.
Hence, since $\pre{t_2} \subseteq m_2$, also $m_2 [t_{2} \rangle m_2'$ is firable, where $m_2' = m_2 \ominus \pre{t_{2}} \oplus \post{t_{2}}$, so that
$(\pre{t_1}, \pre{t_{2}}) \in R^\odot$,  $(\post{t_1}, \post{t_{2}}) \in R^\odot$, $l(t_1) = l(t_{2})$ 
and, 
finally, $(m_1', m_2') \in R^\oplus$, as required, which holds because, from the set $q$
of matching pairs for $m_1$ and $m_2$,
we have removed those in $p$ and we have added
those justifying $(\post{t_1}, \post{t_{2}}) \in R^\odot$. 

If $m_2 [t_2 \rangle m_2'$, then we have to use an argument symmetric to the above, where condition 2 is used instead.
Hence, we have proved that conditions 1 and 2 are enough to prove that $R$ is a place bisimulation.
Finally, the complexity of this procedure is analogous to that in Lemma \ref{pl-rel-dec-lem}.
\fine
\end{lemma}

\begin{theorem}\label{dpl-bis-decid-th}{\bf (D-place bisimilarity is decidable)}
Given a P/T net $N = (S, A, T)$, for each pair of markings $m_1$ and $m_2$, it is decidable whether $m_1 \sim_d m_2$.
\proof
Since $S$ is finite, the set of all the place relations over $S \cup \{\theta\}$ is finite as well. Let us list all the place relations as follows: 
$R_1, R_2, \ldots, R_n$.
Therefore, for $i = 1, \ldots, n$, by Lemma \ref{dpl-rel-dec-lem} we can decide whether the place relation $R_i$ is 
a d-place bisimulation and, in such a case,
we can check whether $(m_1, m_2) \in R_i^\odot$ in $O(k^2\sqrt{k})$ time, where $k$ is the size of the largest marking. 
As soon as we have found a d-place bisimulation $R_i$ such that $(m_1, m_2) \in R_i^\odot$,
we stop concluding that $m_1 \sim_d m_2$. If none of the $R_i$ is a d-place bisimulation such that $(m_1, m_2) \in R_i^\odot$, then
we can conclude that $m_1 \nsim_d m_2$. 
\fine
\end{theorem}

Of course, also this procedure is exponential in the size of the net.

%
\section{A Coarser Decidable Equivalence, not Respecting Causality}\label{i-place-sec}
%

Place bisimilarity is not the coarsest equivalence proposed in \cite{AS92}.
In fact, the coarsest equivalence relation proposed in that paper, called {\em i-place bisimilarity}, is as follows.

\begin{definition}\label{def-i-place-bis}{\bf (I-place bisimulation)}
Let $N = (S, A, T)$ be a P/T net. 
An {\em i-place bisimulation} is a relation
$R\subseteq S \times S$ such that if $(m_1, m_2) \in R^\oplus$
then
\begin{itemize}
\item $\forall t_1$ such that  $m_1[t_1\rangle m'_1$, $\exists t_2$ such that $m_2[t_2\rangle m'_2$, 
$l(t_1) = l(t_2)$ and $(m'_1, m'_2) \in R^\oplus$,
\item $\forall t_2$ such that  $m_2[t_2\rangle m'_2$, $\exists t_1$ such that $m_1[t_1\rangle m'_1$, 
$l(t_1) = l(t_2)$, and $(m'_1, m'_2) \in R^\oplus$.
\end{itemize}

Two markings $m_1$ and $m_2$ are  {\em i-place bisimilar}, denoted by
$m_1 \sim_i m_2$, if there exists an i-place bisimulation $R$ such that $(m_1, m_2) \in R^\oplus$.
\fine
\end{definition}

In other words, $R\subseteq S \times S$ is an i-place bisimulation if $R^\oplus$ is an interleaving bisimulation.
Also $\sim_i$ is an equivalence relation and can be proved decidable with a variation of the proof technique
of Section \ref{decid-place-sec}. In fact, it is easy to prove that
$R$ is an i-place bisimulation if and only if the following two finite conditions are satisfied:
\begin{enumerate}
\item $\forall t_1 \in T$, $\forall m$ such that $(\pre{t_1}, m) \in R^\oplus$
	\begin{itemize}
	\item[$(a)$] $\exists t_2$ such that $m[t_2\rangle m'$, with  $l(t_1) = l(t_2)$ and 
	$(\post{t_1}, m') \in R^\oplus$. 
	\end{itemize}
\item $\forall t_2 \in T$, $\forall m$ such that $(m, \pre{t_2}) \in R^\oplus$
	\begin{itemize}
	\item[$(b)$] $\exists t_1$ such that $m[t_1\rangle m'$,
	with $l(t_1) = l(t_2)$ and $(m', \post{t_2}) \in R^\oplus$.
	\end{itemize}
\end{enumerate}
However, i-place bisimilarity is not very satisfactory, because, even if it respects the branching time, it does not respect causality.

\begin{figure}[!t]
\centering
\begin{tikzpicture}[
every place/.style={draw,thick,inner sep=0pt,minimum size=6mm},
every transition/.style={draw,thick,inner sep=0pt,minimum size=4mm},
bend angle=30,
pre/.style={<-,shorten <=1pt,>=stealth,semithick},
post/.style={->,shorten >=1pt,>=stealth,semithick}
]
\def\eofigdist{2.8cm}
\def\eodist{0.3cm}
\def\eodisty{0.5cm}

\node (a) [label=left:$a)\qquad $]{};

\node (p1) [place]  [label=above:$s_1$] {};
\node (t1) [transition] [below =\eodist of p1,label=left:$a\;$] {};
\node (p2) [place] [right =\eodist of p1,label=above:$s_2\;$] {};
\node (t2) [transition] [below =\eodist of p2,label=right:$b\;$] {};

\draw  [->] (p1) to (t1);
\draw  [->] (p2) to (t2);

  \node (b) [right={3.1cm} of a,label=left:$b)\quad$] {};

\node (p3) [place]  [right=\eofigdist of p1, label=above:$s_3$] {};
\node (t3) [transition] [below left =\eodist of p3,label=left:$a\;$] {};
\node (p4) [place]  [right =\eodisty of p3, label=above:$s_4\;$] {};
\node (t4) [transition] [below right =\eodist of p4,label=right:$b\;$] {};
\node (t5) [transition] [below right =\eodist of p3,label=right:$a\;$] {};

\draw  [->] (p3) to (t3);
\draw  [->] (p4) to (t4);
\draw  [->] (p3) to (t5);
\draw  [->, bend left] (p4) to (t5);
\draw  [->, bend left] (t5) to (p4);

\end{tikzpicture}
\caption{Two i-place bisimilar nets with different causal semantics}
\label{net-i-place}
\end{figure}
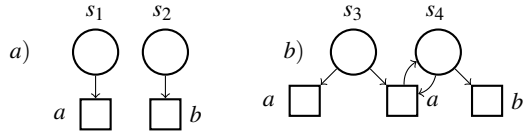

\begin{example}\label{ex-step}
Let us consider the net in Figure \ref{net-i-place}, whose transitions are 
$t_1 = (s_1, a, \theta),$ $ t_2 = (s_2, b, \theta), t_3 = (s_3, a, \theta),$ $t_4 = (s_4, b, \theta)$
and $t_5 = (s_3\oplus s_4, a, s_4)$.
Let us consider relation $R = \{(s_1, s_3), (s_2, s_4)\}$, which is clearly an i-place bisimulation 
(but not a place bisimulation). 
In particular, $(s_1\oplus s_2, s_3\oplus s_4) \in R^\oplus$,
transition $s_3\oplus s_4[t_5\rangle s_4$  is matched by $s_1\oplus s_2 [t_1\rangle s_2$, 
and transition $s_4 [t_4\rangle \theta$ is matched by $s_2 [t_2\rangle \theta$. However, in these 
computations $s_1\oplus s_2$ generates the partial order 
where $a$ and $b$ are independent, while $s_3\oplus s_4$ generates the partial order where
$b$ is caused by $a$.
\fine
\end{example}

Nonetheless, $\sim_i$ implies step bisimilarity $\sim_s$ \cite{NT84}, so that i-place bisimilar markings 
exhibit the same potential parallelism (cf. the subsequent Theorem \ref{id>step-th}).
In fact, markings $s_1 \oplus s_2$ and $s_3 \oplus s_4$ of
the nets in Figure \ref{net-i-place} are the typical example of two markings that are step bisimilar, but not ST 
bisimilar \cite{vGV,GL95,BG02}, a finer behavioral equivalence which, as discussed in \cite{G15}, is the coarsest 
equivalence to be {\em real-time consistent}, meaning that for every association of execution times to actions, 
assuming that actions happen as soon as they can, the running times associated with computations from equivalent 
markings should be the same.
So, i-place bisimilarity,
besides not respecting causality, it is not even real-time consistent, as it does not imply ST bisimilarity. 
(On the contrary,  \cite{G15} argues that
place bisimilarity is real-time consistent, and the same can be argued for d-place bisimilarity.)

Note that i-place bisimilarity is incomparable w.r.t. d-place bisimilarity. On the one hand, Example \ref{ex-dplace},
discussing the net in Figure \ref{net-d-place}, shows that $s_1 \sim_d s_4$, but $s_1 \nsim_i s_4$
because transition $s_1 \deriv{a} s_2$ cannot be matched by $s_4$ as the post-sets $s_2$ and $s_5 \oplus s_6$ have different size.
On the other hand,
Example \ref{ex-step}, discussing the net in Figure \ref{net-i-place}, shows that $s_1 \oplus s_2 \sim_i s_3 \oplus s_4$,
but $s_1 \oplus s_2 \nsim_d s_3 \oplus s_4$ because transition $s_3\oplus s_4 \deriv{a} s_4$ cannot be matched
by $s_1 \oplus s_2$ by preserving the size of the pre-sets.

Of course, it is possible to define an even coarser decidable equivalence by allowing for 
a place to be related to the empty marking $\theta$, as for d-place bisimilarity, yielding i-d-place bisimilarity.

\begin{definition}\label{def-idplace-bis}{\bf (I-d-place bisimulation)}
Let $N = (S, A, T)$ be a P/T net. 
An {\em i-d-place bisimulation} is a relation
$R\subseteq (S\cup \{\theta\}) \times (S \cup \{\theta\})$ such that if $(m_1, m_2) \in R^\odot$
then
\begin{itemize}
\item $\forall t_1$ such that  $m_1[t_1\rangle m'_1$, $\exists t_2$ such that $m_2[t_2\rangle m'_2$, 
$l(t_1) = l(t_2)$ and $(m'_1, m'_2) \in R^\odot$,
\item $\forall t_2$ such that  $m_2[t_2\rangle m'_2$, $\exists t_1$ such that $m_1[t_1\rangle m'_1$, 
$l(t_1) = l(t_2)$ and $(m'_1, m'_2) \in R^\odot$.
\end{itemize}

Two markings $m_1$ and $m_2$ are  {\em i-d-place bisimilar}, denoted by
$m_1 \sim_{id} m_2$, if there exists an i-d-place bisimulation $R$ such that $(m_1, m_2) \in R^\odot$.
\fine
\end{definition}

In other words, $R\subseteq (S\cup \{\theta\}) \times (S \cup \{\theta\})$ is an i-d-place bisimulation if $R^\odot$ is 
an interleaving bisimulation.
Of course, also i-d-place bisimilarity $\sim_{id}$ is an equivalence relation, slightly coarser than $\sim_i$, 
and can be proved decidable with yet another variant
of the proof technique in Sections \ref{decid-place-sec} and \ref{d-place-dec-sec}. Indeed, it is easy to see that a relation $R$ is an i-d-place bisimulation if and only if 
the following two finite conditions are satisfied:
\begin{enumerate}
\item $\forall t_1 \in T$, $\forall m$ such that $(\pre{t_1}, m) \in R^\odot$
	\begin{itemize}
	\item[$(a)$] $\exists t_2$ such that $m[t_2\rangle m'$, with  $l(t_1) = l(t_2)$ and 
	$(\post{t_1}, m') \in R^\odot$. 
	\end{itemize}
\item $\forall t_2 \in T$, $\forall m$ such that $(m, \pre{t_2}) \in R^\odot$
	\begin{itemize}
	\item[$(b)$] $\exists t_1$ such that $m[t_1\rangle m'$,
	with $l(t_1) = l(t_2)$ and $(m', \post{t_2}) \in R^\odot$.
	\end{itemize}
\end{enumerate}

To conclude, we prove that $\sim_{id}$ implies step 
bisimilarity $\sim_s$, so that, even if causality is not respected, at least the concurrent behavior is preserved.

\begin{theorem}\label{id>step-th}
For each P/T net $N = (S, A, T)$, if $m_1 \sim_{id} m_2$ then $m_1 \sim_{s} m_2$.
\proof
If $m_1 \sim_{id} m_2$, then there exists an i-d-place bisimulation $R$ such that $(m_1, m_2) \in R^\odot$. We want to prove that $R^\odot$ is
a step bisimulation, i.e., that 
if $(m_1, m_2) \in R^\odot$
then
\begin{itemize}
\item $\forall G_1$ s.t.  $m_1[G_1\rangle m'_1$, $\exists G_2$ s.t. $m_2[G_2\rangle m'_2$ 
with $l(G_1) = l(G_2)$ and $(m'_1, m'_2) \in R^\odot$,
\item $\forall G_2$ s.t.  $m_2[G_2\rangle m'_2$, $\exists G_1$ s.t. $m_1[G_1\rangle m'_1$ 
with $l(G_1) = l(G_2)$ and $(m'_1, m'_2) \in R^\odot$.
\end{itemize}
To make the presentation lighter, we assume that $G_1$ is composed of two transitions only, say $\{t_1, t_2\}$. The general case when the size 
of $G_1$ is $n \in \nat$ is only notationally more complex, but not more difficult conceptually.

Assume $m_1[\{t_1,t_2\}\rangle m'_1$. This means that $\pre{t_1} \oplus \pre{t_2} \subseteq m_1$. Hence, $m_1 = \pre{t_1} \oplus \pre{t_2} \oplus \overline{m}_1$ and $m_1' = \post{t_1} \oplus \post{t_2} \oplus \overline{m}_1$. 
Since $(m_1, m_2) \in R^\odot$, we may find $m_2^1, m_2^2, \overline{m}_2$ such that $m_2 = m_2^1 \oplus m_2^2 \oplus 
\overline{m}_2$ and, moreover, that $(\pre{t_1}, m_2^1) \in R^\odot$, $(\pre{t_2}, m_2^2) \in R^\odot$ and $(\overline{m}_1, \overline{m}_2) \in R^\odot$.
Since $R$ is an i-d-place bisimulation, from $(\pre{t_1}, m_2^1) \in R^\odot$ it follows that
there exists $t_1'$ such that $m_2^1[t_1'\rangle \overline{m}_2^1$, with  $l(t_1) = l(t_1')$ and 
$(\post{t_1}, \overline{m}_2^1) \in R^\odot$; similarly, from $(\pre{t_2}, m_2^2) \in R^\odot$ it follows that
there exists $t_2'$ such that $m_2^2[t_2'\rangle \overline{m}_2^2$, with  $l(t_2) = l(t_2')$ and 
$(\post{t_2}, \overline{m}_2^2) \in R^\odot$. Therefore, $m_2[\{t_1',t_2'\}\rangle m_2' =  \overline{m}_2^1 \oplus  \overline{m}_2^2 \oplus \overline{m}_2$,
such that $(m_1', m_2') \in R^\odot$ follows by additivity. 

The case when $m_2$ moves first is symmetric, hence omitted. Therefore, $R^\odot$ is a step bisimulation.
\fine
\end{theorem}


%
\section{Conclusion and Future Research}\label{conc-sec}
%

Place bisimilarity seems a very natural behavioral equivalence for finite P/T Petri nets, 
which respects the expected causal behavior,
as it is slightly finer than {\em causal-net bisimilarity} \cite{G15,Gor22}, 
(or, equivalently, {\em structure preserving bisimilarity} \cite{G15}), 
in turn
slightly finer than {\em fully-concurrent bisimilarity} \cite{BDKP91}.
Significantly, to date, it is the first {\em sensible} (i.e., fully respecting causality and branching time) 
behavioral equivalence which has been proved 
decidable for finite P/T nets (with the relevant exception of net isomorphism), even if its 
complexity is exponential in the size of the net.
Thus, it is the only equivalence for which it is possible (at least, in principle) 
to verify algorithmically the (causality-preserving) correctness of a distributed system by 
exhibiting a place bisimulation between its
specification Petri net and its implementation Petri net.

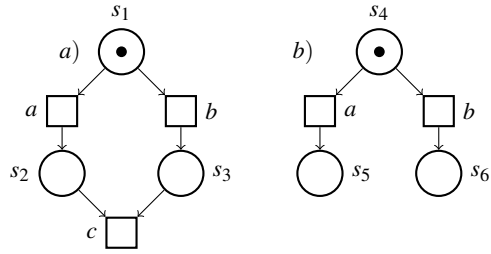
\begin{figure}[t]
\centering
\begin{tikzpicture}[
every place/.style={draw,thick,inner sep=0pt,minimum size=6mm},
every transition/.style={draw,thick,inner sep=0pt,minimum size=4mm},
bend angle=45,
pre/.style={<-,shorten <=1pt,>=stealth,semithick},
post/.style={->,shorten >=1pt,>=stealth,semithick}
]
\def\eofigdist{2.8cm}
\def\eodist{0.3}
\def\eodisty{0.5}

\node (a) [label=left:$a)\quad $]{};

\node (q1) [place,tokens=1] [label={above:$s_1$} ] {};
\node (f1)[transition][below left=\eodisty of q1,label=left:$a$]{};
\node (f2)[transition][below right=\eodisty of q1,label=right:$b$]{};
\node (q2) [place] [below=\eodist of f1,label={left:$s_2$}] {};
\node (q3) [place] [below=\eodist of f2,label={right:$s_3$}] {};
\node (f3)[transition][below right=\eodisty of q2,label=left:$c$]{};

\draw  [->] (q1) to (f1);
\draw  [->] (q1) to (f2);
\draw  [->] (f1) to (q2);
\draw  [->] (f2) to (q3);
\draw  [->] (q2) to (f3);
\draw  [->] (q3) to (f3);


\node (b) [right={2.7cm} of a,label=left:$b)\;\;$] {};

\node (p1) [place,tokens=1]  [right=\eofigdist of q1,label=above:$s_4$] {};
\node (s1) [transition] [below left=\eodisty of p1,label=right:$a$] {};
\node (s2) [transition] [below right=\eodisty of p1,label=right:$b$] {};
\node (p2) [place] [below =\eodist of s1,label= right:$s_5$]{};
\node (p3) [place] [below=\eodist of s2,label= right:$s_6$]{};

\draw  [->] (p1) to (s1);
\draw  [->] (p1) to (s2);
\draw  [->] (s1) to (p2);
\draw  [->] (s2) to (p3);

\end{tikzpicture}
\caption{Two non-place bisimilar nets}
\label{abs-net}
\end{figure}

It is sometimes argued that place bisimilarity is too discriminating. In particular, \cite{ABS91} and \cite{G15} argue
that a {\em sensible} equivalence should not distinguish markings whose behaviors are patently the same, such as
marked Petri nets that differ only in their unreachable parts. As an example, consider the net in Figure \ref{abs-net}, 
originally presented in \cite{ABS91}.
Clearly, markings $s_1$ and $s_4$ are equivalent, also according to all the behavioral equivalences 
discussed in \cite{G15}, except
for place bisimilarity. As a matter of fact, a place bisimulation $R$ containing the pair $(s_1, s_4)$ would require also the pairs
$(s_2, s_5)$ and $(s_3, s_6)$, but then this place relation $R$ cannot be a place bisimulation because 
$(s_2 \oplus s_3, s_5 \oplus s_6) \in R^\oplus$, but $s_2 \oplus s_3$ can perform $c$, while this is not 
possible for $s_5 \oplus s_6$.
Nonetheless, we would like to argue in favor of place bisimilarity, despite this apparent paradoxical example.

As a matter of fact, our interpretation of place bisimilarity is that this equivalence is 
an attempt of giving semantics to {\em unmarked} nets, rather than to marked nets,
so that the focus shifts from the common (but usually undecidable) question 
{\em When are two markings equivalent?} to the more 
restrictive (but decidable) question {\em When are two places equivalent?}
A possible (preliminary, but not accurate enough) answer to the latter question may be: two places are equivalent if, 
whenever the same number of tokens are put on these two places,
the behavior of the marked nets is the same. If we reinterpret the example of 
Figure \ref{abs-net} in this perspective, we clearly see that
place $s_1$ and place $s_4$ cannot be considered as equivalent because, even if the markings $s_1$ and $s_4$
are equivalent, nonetheless the marking $2 \cdot s_1$ is not equivalent
to the marking $2 \cdot s_4$, as only the former can perform the trace $abc$.

A place bisimulation $R$ considers two places $s_1$ and $s_2$ as equivalent if $(s_1, s_2) \in R$, as,  
by definition of place bisimulation, 
they must behave the same in any $R$-related context. Back to our example in Figure \ref{abs-net}, if $(s_1, s_4)$ would 
belong to $R$, then also $(2 \cdot s_1, 2 \cdot s_4)$ should belong to 
$R^\oplus$, but then we discover that the place bisimulation game does not hold for this pair of 
markings, so that $R$ cannot be a place bisimulation.

Moreover, there are at least the following three important technical differences between 
place bisimilarity and other coarser, causality-respecting equivalences, such as fully-concurrent bisimilarity \cite{BDKP91}.
\begin{enumerate}
    \item A fully-concurrent bisimulation is a complex relation --
        composed of cumbersome triples of the form (process, bijection, process)  --
        that must contain infinitely many triples if the net system offers never-ending behavior. 
        (Indeed, not even one single case study of a system with never-ending behavior
        has been developed for this equivalence.)
        On the contrary, a place bisimulation is always a very simple finite relation over the set of places. (And a 
        tiny case study is outlined in Example \ref{ex-pc}.)        
    \item A fully-concurrent bisimulation  proving that $m_1$ and $m_2$ are equivalent
            is a relation specifically designed for the initial markings $m_1$ and $m_2$. If we want to prove that,
            e.g., $n \cdot m_1$ and $n \cdot m_2$ are fully-concurrent bisimilar (which 
            may not hold!), we have to construct a new fully-concurrent bisimulation to this aim. 
            Instead, a place bisimulation $R$ 
            relates those places which are considered equivalent under all the possible 
            $R$-related contexts.  
            Hence, if $R$ justifies that  $m_1 \sim_{p} m_2 $ 
            as $(m_1, m_2) \in R^\oplus$, then
            for sure the same $R$ justifies that $n \cdot m_1$ and $n \cdot m_2$ are 
            place bisimilar, as also 
            $(n \cdot m_1, n \cdot m_2) \in R^\oplus$.            
    \item Finally, while place bisimilarity is decidable, fully-concurrent bisimilarity is undecidable on finite P/T 
    nets with at least two unbounded places \cite{Esp98}. 
\end{enumerate}

Incidentally, note that place bisimilarity $\sim_p$ is not additive (i.e., it is not a congruence w.r.t. multiset union), despite the fact that each place bisimulation is so (as mentioned above in item 2). 
In fact, consider the simple net in Figure \ref{net-tau1} discussed in Example \ref{primo-tau-ex}.
As $(s_1, s_3) \in R_3$, we have that $s_1 \sim_p s_3$; similarly, as $(s_2, s_3) \in R_4$, we have $s_2 \sim_p s_3$; however, 
$s_1 \oplus s_2 \nsim_p 2 \cdot s_3$ because $s_1 \oplus s_2$ can move, while $2 \cdot s_3$ is stuck. 
Additivity (i.e., congruence w.r.t. multiset union) is a very strong property that does not hold for any of the behavioral equivalences discussed in this paper. 
However, \cite{LBJ22} studies the problem of characterizing
the largest  congruence w.r.t. multiset union refining interleaving bisimulation $\sim_{int}$ such that it is also a bisimulation; 
the resulting behavioral relation, called {\em resource bisimilarity}, is proved decidable, even if its complexity is not studied.

One further criticism against place bisimilarity $\sim_p$ may be that its complexity is too high to be useful in practice. However,
on the one hand, we would like to observe that essentially all the decidable properties over unbounded finite P/T nets are at least exponential;
for instance, about the reachability problem (i.e., deciding whether a given marking is reachable from 
the initial marking), Lipton proved an exponential space 
lower bound (EXPSPACE-hard) \cite{Lip}, and none of the algorithms known so far is
primitive recursive (cf., e.g., \cite{Mayr84}). 
On the other hand, we conjecture that our decision procedure can be improved. 
In fact, future work will be devoted to find more efficient algorithms for checking place bisimilarity. One idea could be to
build directly the set of maximal place bisimulations \textendash \, starting from the unique maximal 
equivalence place bisimulation (that can be easily computed
\cite{AS92}) and then elaborating on it, as we did in Example \ref{primo-tau-ex} and \ref{ex-dec} \textendash \, rather than to scan all the 
place relations to check whether
they are place bisimulations, as we did in the proof of Theorem \ref{pl-bis-decid-th}.
Moreover, if the goal is not that of computing the equivalence relation $\sim_p$, but simply to prove that 
two specific markings, say $m_1$ and $m_2$, are place bisimilar, one can simply try to build one particular
place bisimulation $R$ such that
$(m_1, m_2) \in R^\oplus$ (as we did for the tiny case studies of Example \ref{ex-pc} and \ref{ex-bpp}), which seems a much simpler task.

We have proposed a slight weakening of place bisimilarity $\sim_p$, called d-place bisimilarity $\sim_d$, which may 
relate places to the empty marking $\theta$
and which is still decidable. Actually, we conjecture that d-place bisimilarity is the coarsest, 
sensible equivalence relation which is decidable on finite P/T nets (without silent moves). D-place bisimilarity 
is still slightly finer than
fully-concurrent bisimilarity. 

On the subclass of BPP nets 
(i.e., nets whose transitions have singleton pre-set), place bisimilarity coincides with 
{\em team bisimilarity} \cite{Gor17b}, which is actually coinductive and, for this reason, efficiently decidable 
(in polynomial time, by using a generalization of, e.g., the Kanellakis-Smolka's
algorithm \cite{KS83}), characterized by a simple modal logic, 
and even axiomatizable over the process algebra BPP with guarded summation and guarded recursion \cite{Gor17}.
Team bisimilarity is unquestionably the most appropriate behavioral equivalence for BPP nets,
as it coincides with {\em structure-preserving bisimilarity} \cite{G15}, hence matching all the relevant 
criteria expressed in \cite{G15} for 
a sensible behavioral equivalence.
Moreover, d-place bisimilarity specializes to {\em h-team} bisimilarity \cite{Gor22}, which coincides
with fully-concurrent bisimilarity on this subclass of P/T nets.

We have also discussed that place bisimilarity is not the coarsest decidable equivalence 
for Petri nets proposed in \cite{AS92}.
In fact, we have proved that i-place bisimilarity  $\sim_i$ is still decidable, even if not satisfactory 
as it does not respect causality. For the sake of exposition completeness,
we have observed that the newly proposed variant i-d-place bisimilarity $\sim_{id}$ is an even coarser decidable equivalence relation, that 
is the coarsest
decidable equivalence defined so far on finite Petri nets respecting the branching time and the concurrent behavior (as it is finer than step bisimilarity $\sim_s$), but not respecting causality.

Extensions of the theory presented in this paper have been sketched in \cite{Gor-forte21}, where the decidable 
{\em branching place bisimilarity} has been proposed for finite
P/T nets with silent moves (taking inspiration from branching bisimilarity on LTSs \cite{vGW96}),
and also in \cite{CG-mfcs21}, where the decidable {\em pti-place bisimilarity} has been proposed for the
Turing-complete model of finite Petri nets with inhibitor arcs \cite{FA73,Pet81}.
Future work will be devoted to deepen the study of these extensions and to experimentally show the usefulness of the 
place bisimulation framework over some complex case studies.


\begin{thebibliography}{11}





\bibitem{FA73}
T. Agerwala, M. Flynn,
\newblock Comments on capabilities, limitations and correctness of Petri nets,
\newblock {\em SIGARCH Comput. Archit. News}, 2(4):81-86, 1973.

\bibitem{ABS91}
C. Autant, Z. Belmesk, Ph. Schnoebelen,
\newblock  Strong bisimilarity on nets revisited, 
\newblock in Procs. PARLE'91, LNCS 506, 295-312, Springer, 1991.


\bibitem{AS92}
C. Autant, Ph. Schnoebelen,
\newblock  Place bisimulations in Petri nets, 
\newblock in Procs. Application and Theory of Petri Nets 1992, LNCS 616, 45-61, Springer, 1992.

\bibitem{BD87}
E. Best, R. Devillers, 
\newblock  Sequential and concurrent behavior in Petri net theory,
\newblock {\em Theoretical Computer Science} 55(1):87-136, 1987.


\bibitem{BDKP91}
E. Best, R. Devillers, A. Kiehn, L. Pomello,
\newblock  Concurrent bisimulations in Petri nets,
\newblock {\em Acta Inf.}  28(3): 231-264, 1991.

%
\bibitem{BG02}
M. Bravetti, R. Gorrieri,
\newblock  Deciding and axiomatizing weak ST bisimulation for a process algebra with recursion and action refinement,
\newblock {\em ACM Trans. Comput. Log.} 3(4): 465-520, 2002.


\bibitem{CG-mfcs21}
A. Cesco, R. Gorrieri,
\newblock A decidable equivalence for a Turing-complete, distributed model of computation,
\newblock in Procs. {\em 46$^{th}$ International Symposium on Mathematical Foundations of Computer Science} (MFCS'21), 
28:1-28:18,  LIPIcs 202, Schloss Dagstuhl - Leibniz-Zentrum f\"ur Informatik, 2021.

%
\bibitem{DDM89}
P.~Degano, R.~De~Nicola, U.~Montanari,
\newblock Partial ordering descriptions and observations of nondeterministic concurrent systems,
\newblock in (J. W. de Bakker, W. P. de Roever, G. Rozenberg, Eds.)
\newblock {\em Linear Time, Branching Time and Partial Order in Logics and Models for Concurrency}, LNCS 354, 438-466, Springer, 1989.


\bibitem{DesRei98}
J. Desel, W. Reisig,
\newblock Place/Transition Petri nets,
\newblock  in {\em Lectures on Petri Nets I: Basic Models}, 
LNCS 1491, 122-173, Springer, 1998.



\bibitem{Esp98}
J. Esparza,
\newblock Decidability and complexity of Petri net problems: An introduction,
\newblock {\em Lectures on Petri Nets I: Basic Models},
\newblock   LNCS 1491, 374-428, Springer, 1998.

%
\bibitem{vGV}
R.J.~van Glabbeek, F.~Vaandrager,
\newblock Petri Net models for algebraic theories of concurrency,
\newblock in Proc. {\em PARLE'87}, LNCS 259,  224-242, Springer-Verlag, 1987.

%
\bibitem{vGG89}
R.J. van Glabbeek, U. Goltz,
\newblock  Equivalence notions for concurrent systems and refinement of actions,
\newblock in Procs. MFCS'89, LNCS 379, 237-248, Springer, 1989.

\bibitem{vGW96}
R.J. van Glabbeek, W.P. Weijland,
\newblock  Branching time and abstraction in bisimulation semantics,
\newblock {\em Journal of the ACM} 43(3):555-600, 1996.



\bibitem{G15}
R.J. van Glabbeek,
\newblock  Structure preserving bisimilarity - Supporting an operational Petri net semantics of CCSP, 
\newblock in (R. Meyer, A. Platzer, H. Wehrheim, Eds.)
\newblock {\em Correct System Design} --- Symposium in Honor of Ernst-R\"udiger Olderog on the Occasion of His 60th Birthday,  
LNCS 9360, 99-130, Springer, 2015.

\bibitem{GR83}
U. Goltz, W. Reisig,
\newblock The non-sequential behaviour of Petri nets,
\newblock {\em Information and Control} 57(2-3):125-147, 1983.





\bibitem{Gor17}
R.~Gorrieri,
\newblock {\em Process Algebras for Petri Nets: The Alphabetization of Distributed Systems},  
\newblock EATCS Monographs in Computer Science, Springer, 2017.


\bibitem{Gor17b}
R.~Gorrieri,
\newblock Team bisimilarity, and its associated modal logic, for BPP nets, 
\newblock {\em Acta Informatica} 58(5):529-569, 2021. 


\bibitem{Gor20c}
R.~Gorrieri,
\newblock Causal semantics for BPP nets with silent moves,
\newblock {\em Fundamenta Informaticae}, 180(3):179-249, 2021. 

\bibitem{Gor22}
R.~Gorrieri,
\newblock A study on team bisimulation and h-team bisimulation for BPP nets, 
{\em Theoretical Computer Science} 897:83-113, 2022.

\bibitem{Gor-forte21}
R. Gorrieri,
\newblock Branching place bisimilarity: A decidable behavioral equivalence for finite Petri nets with silent moves,
\newblock in Procs. {\em 41$^{st}$ Formal Techniques for Distributed Objects, Components, and Systems} (FORTE'21), 
LNCS 12719, 80-99, Springer, 2021.

%
\bibitem{GL95}
R. Gorrieri, C. Laneve,
Split and ST bisimulation semantics,
{\em Inf. Comput.} 118(2): 272-288, 1995.


\bibitem{HK73}
J.E. Hopcroft, R.M. Karp,
\newblock  An $n^{5/2}$ algorithm for maximum matchings in bipartite graphs, 
\newblock {\em SIAM Journal on Computing}, 2 (4): 225-231,1973.


\bibitem{Jan95}
P. Jan\u{c}ar,
\newblock Undecidability of bisimilarity for Petri nets and some related problems,
\newblock {\em Theoretical Computer Science} 148(2):281-301, 1995.


%
\bibitem{KS83}
P. Kanellakis, S. Smolka, 
\newblock CCS expressions, finite state processes,
and three problems of equivalence, 
\newblock in Procs. 2nd Annual ACM Symposium on
Principles of Distributed Computing, 228-240, ACM Press, 1983.


\bibitem{Kar73}
A. V. Karzanov, An exact estimate of an algorithm for finding a maximum flow, applied to the problem on representatives, 
{\em Problems in Cybernetics} 5: 66-70, 1973. 


\bibitem{Kel76}
R. Keller,
\newblock Formal verification of parallel programs,
\newblock {\em Comm. of the ACM} 19(7):561-572, 1976.




%
\bibitem{Lip}
R. Lipton,
\newblock The reachability problem requires exponential space,
\newblock Tech. Rep. 63, Yale University, 1976.


\bibitem{LBJ22}
I.A. Lomazova, V.A. Bashkin, P. Jan\u{c}ar,
\newblock Resource bisimilarity in Petri nets is decidable,
\newblock {\em Fundamenta Informaticae}, 186(1-4):175-194, 2022. 


%
\bibitem{Mayr84}
E.W. Mayr,
\newblock An algorithm for the general Petri net reachability problem,
\newblock {\em SIAM J. Comput.} 13:441-460, 1984.

\bibitem{Mil89} R. Milner. {\it Communication and Concurrency},
Prentice-Hall, 1989.

%
\bibitem{NT84}
M. Nielsen, P.S. Thiagarajan,
\newblock  Degrees of non-determinism and concurrency: A Petri net view,
\newblock in Procs. of the Fourth Conference on Foundations of Software Technology and Theoretical Computer Science (FSTTCS'84),
LNCS 181, 89-117, Springer-Verlag, 1984.


\bibitem{Old}
E.R.~Olderog,
\newblock {\em Nets, Terms and Formulas},
\newblock Cambridge Tracts in Theoretical Computer Science 23, Cambridge University Press, 1991.


\bibitem{PT87}
R. Paige, R.E. Tarjan,
\newblock  Three partition refinement algorithms,
\newblock {\em SIAM Journal of Computing} 16(6):973-989, 1987.


\bibitem{Park81}
D.M.R. Park,
\newblock Concurrency and automata on infinite sequences,
\newblock In Proc. 5th GI-Conference on Theoretical Computer Science, LNCS 104, 167-183, 
Springer, 1981.



\bibitem{Pet81}
J.L. Peterson,
\newblock {\em Petri Net Theory and the Modeling of Systems}, Prentice-Hall, 1981.

\bibitem{RT88}
A. Rabinovich,  B.A. Trakhtenbrot,
\newblock Behavior structures and nets,
\newblock {\em Fundamenta Informaticae} 11(4):357-404, 1988.


\bibitem{Reisig}
W. Reisig,
\newblock {\em Petri Nets: An Introduction},
\newblock EATCS Monographs in Theor. Comp. Science,
\newblock Springer, 1985.

%
\bibitem{Rei98}
W. Reisig,
\newblock {\em Elements of Distributed Algorithms: Modeling and Analysis with Petri Nets},
\newblock Springer-Verlag, 1998.

%
\bibitem{RR98b}
W. Reisig, G. Rozenberg (eds.),
\newblock {\em Lectures on Petri Nets II: Applications},
\newblock   Lecture Notes in Computer Science 1492,
\newblock Springer-Verlag, 1998.



\end{thebibliography}
\end{document}